\documentclass[a4paper, 11pt]{article}
\usepackage[T1]{fontenc} 
\usepackage[utf8]{inputenc}
\usepackage[english]{babel}
\usepackage{amssymb,amsmath,amsthm}
\usepackage{booktabs}
\usepackage{subcaption}
\usepackage{authblk}
\usepackage{hepnames}
\usepackage{siunitx}
\usepackage{graphicx}
\usepackage{csquotes}
\usepackage{array}
\usepackage{comment}
\usepackage[top=1.25in, bottom=1.75in,
left=1.25in, right=1.25in]{geometry}

\usepackage[backend=biber,sorting=none]{biblatex}
\bibliography{biblio.bib}

\newcommand{\df}[1]{\ensuremath{\operatorname{d}\!{#1}}}

\newcommand{\newsdm}{NEWSdm}
\newcommand{\geant}{GEANT4}
\DeclareSIUnit\erg{erg}
\DeclareSIUnit\parsec{pc}
\DeclareSIUnit\ton{ton}
\DeclareSIUnit\year{y}

\usepackage{hyperref}
\hypersetup
{
    pdfauthor = {G.\ De Lellis,
    A.\ Di Crescenzo, A.\ Gallo Rosso, V.\ Gentile
    F.\ Vissani},
    pdfsubject={Paper NEWSdm 04/2018},
    pdftitle={Supernova neutrino physics with
    a nuclear emulsion detector},
    pdfcreator={\LaTeX},
    pdfkeywords={dark matter, supernova neutrinos,
    nuclear emulsion},
    hidelinks,
}

\title{Supernova neutrino physics with
    a nuclear\\ emulsion detector}
    
\author[1,2]{G.\ De Lellis}
\author[1,2]{A.\ Di Crescenzo}
\author[3,4]{A.\ Gallo Rosso\thanks{Corresponding
author: \href{mailto:andrea.gallorosso@gssi.it}
{andrea.gallorosso@gssi.it}.}}
\author[3,4]{V.\ Gentile\thanks{Corresponding
author: \href{mailto:valerio.gentile@gssi.it}
{valerio.gentile@gssi.it}.}}
\author[4]{F.\ Vissani}
\affil[1]{Università Federico II di Napoli,
	Dipartimento di Fisica,\protect\\
	Via Cintia 21, 80126 Napoli, Italy}
\affil[2]{INFN, Sezione di Napoli, Via Cintia 21, 80126 Napoli, Italy}
\affil[3]{Gran Sasso Science Institute,
	Viale F.\ Crispi 7, 67100 L'Aquila, Italy}
\affil[4]{INFN,
	Laboratori Nazionali del Gran Sasso,
	Via G. Acitelli 22,\protect\\
	67100 Assergi (AQ), Italy\vspace{-3em}}
\date{}                     
\begin{document}
	\maketitle
\begin{abstract}
 The existence of the coherent neutrino-nucleus scattering reaction requires to evaluate, for any detector devoted to WIMP searches, the irreducible background due to conventional neutrino sources and at same time, it gives a unique chance to reveal supernova neutrinos. We report here a detailed study concerning a new directional detector, based on the nuclear emulsion technology. A Likelihood Ratio test shows that, in the first years of operations and with a detector mass of several tens of tons, the observation of the supernova signal
can be achieved. The determination of the distance of the supernova from the neutrinos and the observation of \textsuperscript{8}B neutrinos are also discussed.	
\end{abstract}
  
\section*{Introduction}
It has been estimated that, at a redshift
$z\simeq 0.4$, the rate of core-collapse supernovae
is about $10^6$ per year \cite{Jorgensen97}.
However, when we restrict ourselves to  the subset of those that are
sufficiently close to produce a detectable neutrino signal,
the rate is just a few per century, see e.g.\ \cite{Vissani:ConferenzaSN}.
This makes the detection of neutrinos from the next galactic
core-collapse supernova a once-in-a-lifetime
opportunity of taking a glimpse into extreme
physics, such as the formation of neutron stars (or other
compact remnants),  explosion of massive stars,
correlation with other types of radiation,
as well as standard or non-standard
neutrino oscillations mechanisms \cite{Volpe:2016bkp}.

Standard supernova detectors such as Super-Kamiokande
\cite{Fukuda:2002uc}, LVD \cite{Agafonova:2014leu}
or IceCube \cite{Bruijn:2013ibl} are constantly ready
to reveal the next galactic supernova explosion 
and in case of an
event they will provide us with huge data sets, although consisting
mainly of electron antineutrinos.
As pointed out in \cite{GalloRosso:2017hbp, GalloRosso:2017mdz}, it is essential to observe other types of neutrinos, to achieve the minimal goal of
measuring the total energy emitted in neutrinos with a reasonable accuracy.

More in general, in order to fully profit  of the scientific occasion offered by a future galactic supernova explosion and
to try and obtain a complete picture of supernova neutrino emission, it will be necessary to observe
as many channels as possible, including a significant neutral-current sample.
We recall that neutral-current
events are free from the uncertainties that currently affect
the description of neutrino oscillations.

Detectors aiming at WIMP dark matter detection
have to reach high standards of purity and noise reduction,
which makes them also good supernova
neutrino detectors,
via coherent elastic neutrino scattering on
nuclei. In fact, the nuclear recoil on silver target (Ag)
is the same (\SI{6}{\kilo\electronvolt}) if given
by a WIMP with mass $m_{\chi}\sim\SI{25}{\giga\electronvolt}$
travelling at $v_{\chi}\sim\SI{260}{\kilo\meter\per\second}$
or by a neutrino with energy
$E_{\nu}\sim\SI{17}{\mega\electronvolt}$.
This consideration already illustrates that
supernova neutrinos and  high
energy solar neutrinos are potentially detectable in suitable
dark matter detectors. The other key requirements of the
detectors are: a sufficient mass, a continuous operation, stability.
In the absence of an external trigger, a smoking gun for
the identification of supernova neutrinos is the
measurement of their incoming direction.
This can be implemented 
thanks to innovative dark matter (and neutrino) detectors based on very-fine grained (nanometric) nuclear emulsion films~\cite{Asada:2017wvp}.

In this paper we have evaluated quantitatively and for the
first time the possibility of
detecting and identifying coherent elastic neutrino nucleus
scattering with nanometric emulsion detectors exploiting directional information.
We will consider the setup 
foreseen for the \newsdm\ detector, whose nanometric nuclear emulsion 
films are capable to reconstruct the trajectories down to
\SI{50}{\nano\meter}
and with an intrinsic angular resolution of about $13^{\circ}$ \cite{Aleksandrov:2016fyr}.

\section{Supernova neutrinos}
\label{sec:model}

Supernova neutrino fluxes
are predicted by means of complex simulations
affected by significant uncertainties,
which limit the chances of reliably predicting the aftermath of
core collapse events, and in particular, the interaction rates.
The distance of the next gravitational collapse in the Milky
Way can be estimated statistically by the study of the
astronomical precursors and descendants of the supernovae of
this type, that have all similar spatial distributions. 
The typical average distance is slightly more than the
distance of the Galactic centre and the variance is
significant; the simplest reasonable estimation 
$\left(10\pm 4.5\right)\si{\kilo\parsec}$
\cite{Costantini:2005un} is rather similar to those of
other more sophisticate evaluations, such as
$\left(10.7\pm 4.9\right)\si{\kilo\parsec}$
\cite{Mirizzi:2006xx} or
$\left(10\pm 4\right)\si{\kilo\parsec}$
from Fig.~2 of \cite{Adams:2013ana}.
However, note that these are mean values; the mode of the distance
distribution is around \SI{8.5}{\kilo\parsec},
i.e.\ the distance from the 
Galactic Centre \cite{Costantini:2005un}.
Since this value is
compatible within the range given in the
former evaluations, in the following we will consider a supernova explosion occurring at a distance of $D = \SI{8}{\kilo\parsec}$
which is the conventional value adopted for this type of calculations.

The formation of a neutron star requires to radiate $\mathcal{E}_{\mathrm{B}}\sim
\SI{3e53}{\erg}$ in neutrinos.
This value is the standard one
usually assumed in literature (see e.g.\
\cite{Lujan-Peschard:2014lta,
Loredo:2001rx,Pagliaroli:2008ur}).
The total energy is assumed to be equally
distributed among the six neutrino species:
\begin{equation}
	\label{eq:etot}
	\mathcal{E}_{i} =
	\SI{0.5e53}{\erg}\quad\text{with}\quad
	i = \Pnue,\,\APnue,\,\nu_{x}.
\end{equation}
The notation $\nu_x$ stands for
one among the four non-electronic species:
$\nu_\mu$, $\bar{\nu}_\mu$, $\nu_\tau$,
$\bar{\nu}_\tau$.
The energy equipartition is a
useful working hypothesis,
even if it might be true only
within a factor two \cite{Raffelt:2005fb}.
Mean energies are expected to follow the hierarchy
$\langle  E_{\Pnue}\rangle \le \langle
E_{\APnue}\rangle
\le \langle  E_{\nu_{x}}\rangle$
but the debate about their
ratios is still open \cite{Raffelt:2005fb}.
In the following we adopt the choice proposed in Ref.~\cite{Lujan-Peschard:2014lta}:
\begin{equation}
	\label{eq:emedie}
	\langle E_{\nu_e}\rangle =
	\SI{9.5}{\mega\electronvolt},\qquad
	\langle E_{\bar{\nu}_e}\rangle
	= \SI{12}{\mega\electronvolt},\qquad
	\langle E_{\nu_x}\rangle
	= \SI{15.6}{\mega\electronvolt}.
\end{equation}
These values are in agreement with numerical simulations
and also with the experimental observations from SN 1987A
\cite{Loredo:2001rx,Pagliaroli:2008ur,Vissani:2014doa}.

We use time integrated fluxes (fluences) derived from
the quasi-thermal parameterisation firstly proposed in Ref.~\cite{Keil:2002in}:\footnote{In the following 
$\hbar = c = k_B = 1$ is assumed.}
\begin{equation}\label{eq:dFdE}
	\mathcal{F}_i\left(E_{\nu}\right) = 
	\frac{\df F_i}{\df E_{\nu}} = 
	\frac{\mathcal{E}_i}{4\pi D^2}
	\frac{E_{\nu}^{\alpha_i}\:
	e^{-E_{\nu}/T_i}}{T_i^{\alpha_i +2 }\:
	\Gamma\left(\alpha_i+2\right)}
	\quad\text{with}\quad
	i = \nu_e,\, \bar{\nu}_e,\, \nu_x
\end{equation}
where $E_{\nu}$ is the neutrino energy, $\Gamma(x)$
is the Euler gamma function and $T_i$ is the temperature,
linked to the mean energy by the relation:
\begin{equation}
	T_i = \langle E_i\rangle / (\alpha_i + 1).
\end{equation}
The so-called 
\textit{pinching parameter} $\alpha$ parametrises the deviation
from the thermal Maxwell-Boltzmann distribution (reproduced
by $\alpha = 2$).

\begin{figure}[t]
\centering
\includegraphics[width=0.7\textwidth]{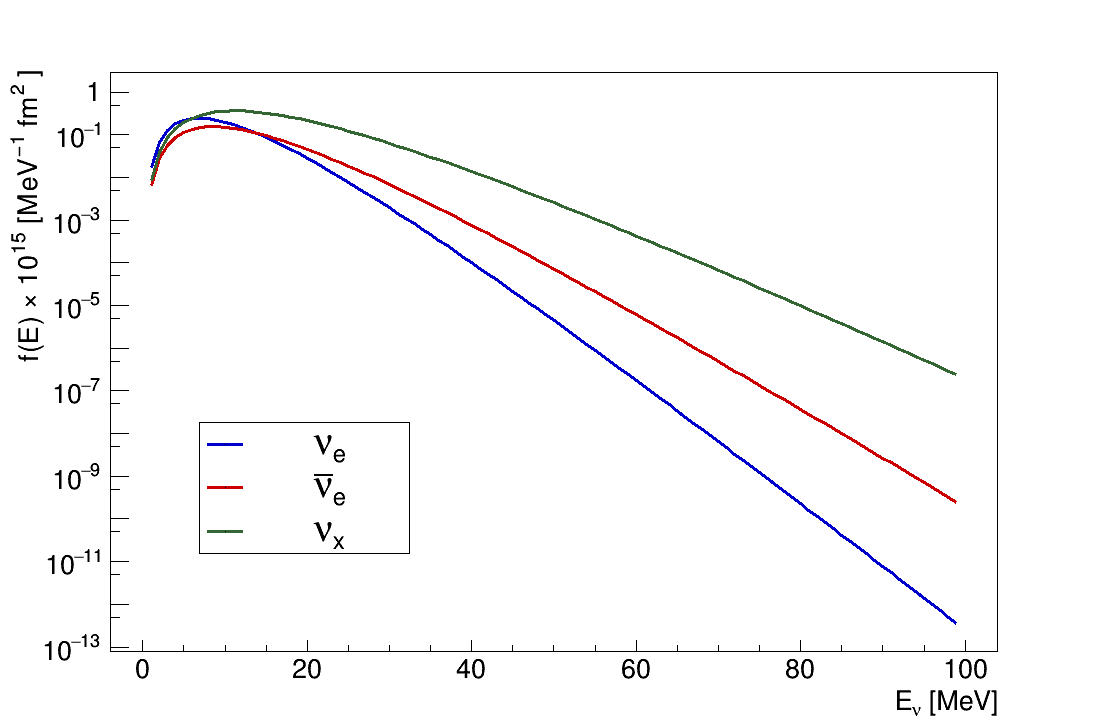}
\caption{Fluences for the three neutrino
    species as defined in \eqref{eq:dFdE} with the 
    parameters indicated in \eqref{eq:etot},
    \eqref{eq:emedie} and \eqref{eq:alfai}.}
\label{fig:fluenza}
\end{figure}

The fluences are expected to be described by distributions
with $\alpha$ slightly larger but
not far from 2, namely, mild deviations from the hypothesis
of a thermal distribution --- see Ref.\
\cite{Vissani:2014doa,GalloRosso:2017hbp,Tamborra:2012ac, GalloRosso:2017mdz}
for a discussion. In the following calculations, we will adopt the value
\begin{equation}\label{eq:alfai}
	\alpha_i = 2.5 \quad\text{with}\quad
	i = \nu_e,\,\bar{\nu}_e,\,\nu_x.
\end{equation}
As discussed in Ref.\ \cite{Vissani:2014doa},
is consistent with the signal seen from SN1987A.
Fluences are reported in Fig.~\ref{fig:fluenza} as a
function of the above defined parameters.

\subsection{Neutral-current coherent interaction}

The differential cross-section in the
solid angle $\Omega$, describing
the coherent interaction between a neutrino and a
nucleus mediated by the \PZzero boson,
can be written as \cite{Freedman:1973yd}:
\begin{equation}\label{eq:dSdO}
	\frac{\df\sigma}{\df\Omega} =
	\frac{G_F^2}{\left(2\pi\right)^2}\frac{Q_w^2}{4}
	\frac{ E_{\nu}^2\left(1+\cos\theta\right)^2}{
	\left[1 + \left(1-\cos\theta\right)\,E_{\nu}/M
	\right]^3}
	F^2\left(Q\right)
\end{equation}
where $G_F$ is the Fermi constant, $E_{\nu}$ is
the neutrino energy, $\theta$ the angle of the
scattered neutrino with respect to the incoming
direction, $M$ is the nucleus mass\footnote{
A mass independent expression is obtained by neglecting the terms order $E_{\nu}/M$.}
and $Q_w$ is the weak charge. For a 
nucleus with $N$ neutrons and $Z$ protons its
value is:
\begin{equation}
	Q_w = N - \left(1 - 4\sin^2\vartheta_W\right)Z.
\end{equation}
with $\vartheta_W$ the Weinberg angle. It
depends on the transferred four-momentum
$Q$, but for
$Q \lesssim \SI{100}{\mega\electronvolt}$
it is almost constant and equal to
\cite{Erler:2004in}:
\begin{equation}
	\sin^2\vartheta_W \approx 0.239.
\end{equation}

\begin{figure}[t]
\centering
\includegraphics[width=0.7\textwidth]{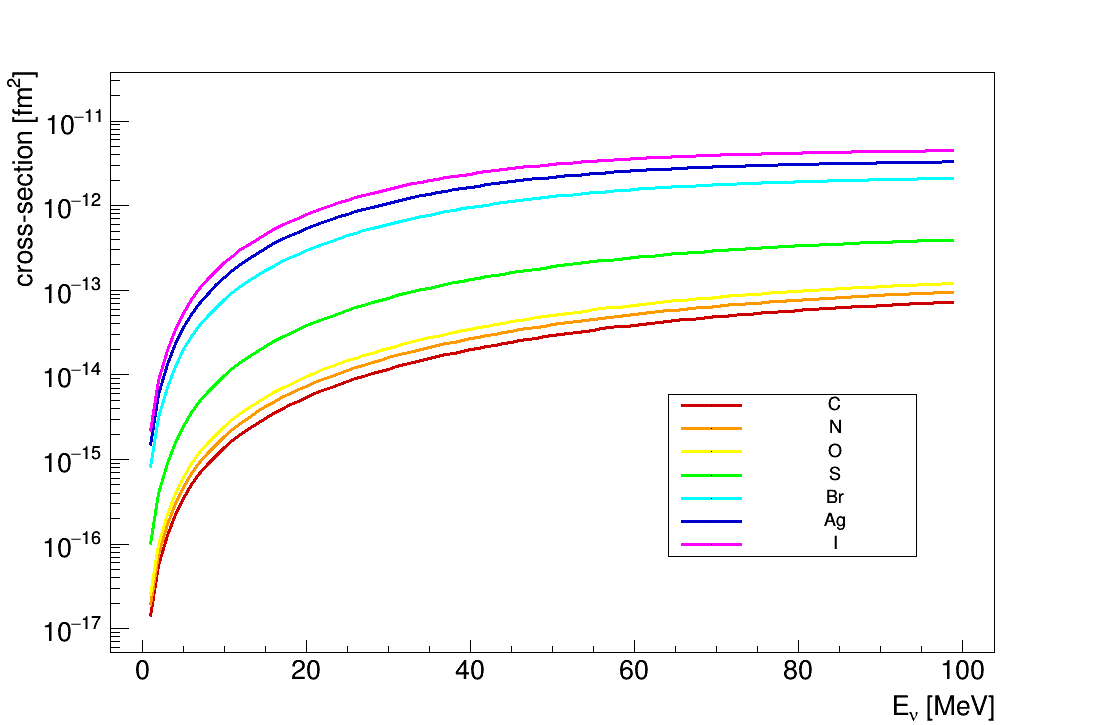}
\caption{
    Energy dependent cross-section for neutral-current
    neutrino-nucleus interaction after integrating the differential
    cross-section \eqref{eq:dSdO} over the solid angle.
}
\label{fig:sezione}
\end{figure}

The function $F\left(Q\right)$ is the form
factor, accounting for
the nucleus size, when
the transferred four-momentum is nonzero.
In plane-waves approximation it is
the Fourier transform of
the nuclear charge distribution. Following Ref.\ 
\cite{Lewin:1995rx} and using the density
given by the Helm model \cite{Helm1956},
one obtains:
\begin{equation}
F(Q) = \frac{3}{Q\,R_n} \left[\frac{\sin(Q\,R_n)}
{(Q\,R_n)^2}-
\frac{\cos(Q\,R_n)}{(Q\,R_n)}\right]\text{exp}
\left(-\frac{(Q\,s)^2}{2}\right),
\end{equation}
where $s \approx 0.9$ fm and:
\begin{equation}
	R^2_n (\mathrm{fm}^2) \approx
	1.5129\,A^{2/3} - 1.4760 A^{1/3} + 2.5371,
\end{equation}
with $A$ the nucleus mass number.
Combining all together, and integrating
\eqref{eq:dSdO} over the solid angle, one
obtains the expression for the
energy dependent total cross-section
$\sigma\left(E_{\nu}\right)$, shown
in Fig.~\ref{fig:sezione} for the
target elements considered in the next sections.

\begin{figure}[t]
	\centering
	\includegraphics[width=0.7\textwidth]{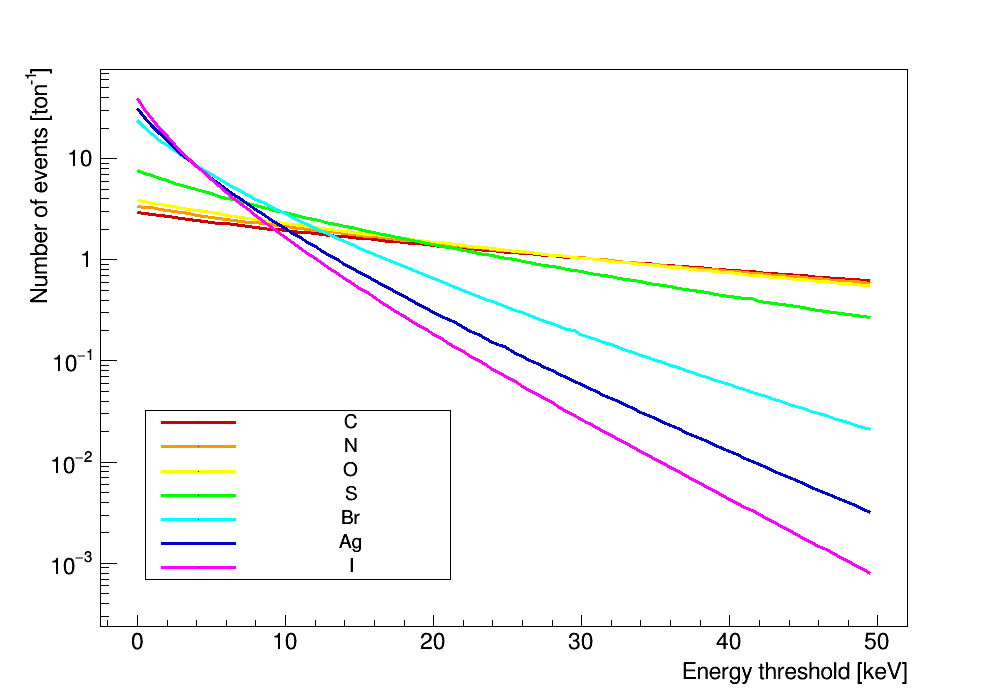}
	\caption{Total number of events
    of supernova neutrino induced recoils
    per ton of active mass, as a function of the threshold
    on the recoiling nucleus energy. Different colours represent 
    different target elements. 
	}
	\label{fig:eventi}
\end{figure}

Following Ref.\ \cite{BIASSONI2012151},
the differential rate of expected events in
the nucleus recoiling energy $K$ can be written
as:
\begin{equation}
	\frac{\df{\mathrm{N}}_{i}}{\df K} =
	N_T \iint \df{\Omega}\df{E_{\nu}}
	\frac{\df\sigma}{\df\Omega}
	\mathcal{F}_i\left(E_{\nu}\right)
	\delta\left(K - \frac{Q^2}{2 M}\right),
\end{equation}
where $M$ is the nucleus mass, $N_T$ the number
of target nuclei and $\delta$ is the Dirac
delta function. Using the expression
\eqref{eq:dSdO} and the properties of
the delta function one obtains:
\begin{equation}\label{eq:dNdK}
	\begin{aligned}
	\frac{\df{\mathrm{N}}_{i}}{\df K} =& 
	N_T \frac{G_F^2}{2\pi}\frac{Q_w^2}{4} F^2\left(
	2 M K\right) \int_{E_{\mathrm{inf}}}^{\infty}
	\df{E_{\nu}}
	\times\\ &\times
	\mathcal{F}_i\left(E_{\nu}\right)
	\left(1 +
	\frac{M K + K E_{\nu} - E_{\nu}^2}
	{\left(K - E_{\nu}\right) E_{\nu}}
	\right)
	\frac{M E_{\nu}^2}{\left(K - E_{\nu}\right)^2},
	\end{aligned}
\end{equation}
with:
\begin{equation}\label{eq:Emin}
	E_{\mathrm{inf}} = \frac{1}{2}\left( K +
	\sqrt{2 M K + K^2}
	\right).
\end{equation}
Integrating  \eqref{eq:dNdK} over the energy from the detector
threshold $K_{\mathrm{thr}}$, one can obtain
the number of interactions of the $i$-th neutrino
species. Summing over all the neutrino species one
obtains the total number of expected events,
shown in Fig.~\ref{fig:eventi} as a function
of the nucleus recoil energy threshold, for a ton
of active mass. The elements considered here are
the same as in Fig.~\ref{fig:sezione}.

\section{\newsdm\ detector}\label{sec:det}

The next generation of ton-scale detectors for dark matter search will be sensitive to neutrinos coming both from the Sun and supernova explosions.
Neutrino sources lead to detectable events the current experiments due
to the coherent elastic neutrino nucleus scattering (CE$\nu$NS). This 
can be a source of {\em background}, unless these events can be tagged by means of directional sensitivity.

Measuring the direction of nuclear recoils appears to be the only strategy to go beyond the neutrino bound.
On top of that, directional dark matter detectors operating during a supernova explosion can identify the {\em signal} due to neutrino interactions. 

The \newsdm\ experiment (Nuclear Emulsions for WIMP Search with directional measurement) \cite{Aleksandrov:2016fyr} is a detector for a directional dark matter search.
Nuclear emulsions are made of crystals of silver halide  immersed in an organic gelatine.
The passage of a charged particle through the emulsion produces along its path atomic-scale perturbations, called latent images. The chemical treatment makes
Ag grains visible with an optical microscope. A track will therefore result in a sequence of aligned grains.

The \newsdm\ detector consists of a bulk of new generation nuclear emulsion films called \emph{Nano Imaging Trackers} (NIT) with nanometric size grains \cite{Natsume:2007zz}. They are produced at Nagoya University in Japan where emulsions with \SI{20}{\nano\meter} of diameter crystals  have been already produced \cite{Asada:2017wvp}. 
The reconstruction of trajectories with path lengths as short as \SI{50}{\nano\meter} is possible if analysed by means of microscopes with
enough resolution. The chemical composition of these emulsion films is reported in Tab.~\ref{tab:element}.

\begin{table}[t]
  \centering
    \begin{tabular}{ccc}
        \toprule
        Element & Mass fraction  & Atomic Fraction \\     
        \midrule 
        Ag      & 0.44           &    0.10         \\
        Br      & 0.32           &    0.10         \\
        C       & 0.101          &    0.214        \\
        O       & 0.074          &    0.118        \\
        N       & 0.027          &    0.049        \\
        I       & 0.019          &    0.004        \\
        H       & 0.016          &    0.410        \\
        S       & 0.003          &    0.003        \\
        \midrule
        \multicolumn{3}{c}{Density \quad 3.43 g/cc} \\
        \bottomrule
        \end{tabular}
    \caption{Chemical composition of NIT emulsions.}
    \label{tab:element}
\end{table}

The \newsdm\ experiment uses fully automated optical microscopes to perform the emulsion analysis. A 3D reconstruction with nanometric accuracy is currently feasible. Moreover, the R\&D for a new scanning system is ongoing to enable the analysis of ton-scale detectors on a time scale comparable with the exposure time \cite{Alexandrov:2015kzs, Alexandrov:2016tyi, 2017Nature}.

The challenging task of detecting track lengths shorter than the diffraction limit --- $\sim \SI{200}{\nano\meter}$ --- is achieved adopting a two-step approach: (i) candidate selection with elliptical shape analysis, (ii)
candidate validation with polarised light analysis.
The scanning with optical microscope cannot distinguish two grains closer than $\sim \SI{200}{\nano\meter}$ since they would appear as a single cluster, according to the Rayleigh criterion for visible light. Nevertheless, the cluster would have an elliptical shape
with the major axis along the actual direction of the recoiled nucleus, unlike single grains from thermal excitation that would appear as spherical \cite{KIMURA201212}.

The candidate tracks selected by the elliptical shape analysis are then validated using an innovative technique which allows to overcome the intrinsic limit of the optical resolution: the
polarised light analysis. 
The resonance effect of polarised light, occurring when nanometric metallic (silver) grains are dispersed in a dielectric medium (organic gelatine), is sensitive to the shape of nanometric
grains \cite{TAMARU2002}: since silver grains in the emulsions are not spherical, the resonant response depends on the polarisation of the incident light.
The shortest detectable tracks are made of two grains, to be distinguished from clusters produced by a single grain.

Given the random orientation of the two grains building up the track, each grain is emphasised at different polarisation values. Taking multiple measurements over the whole polarisation 
range produces a displacement of the barycentre of the cluster, thus becoming sensitive to the actual presence of two grains within the cluster.

On the contrary single grains do not produce any displacement and can therefore be distinguished.
Using this effect, an unprecedented accuracy better than 10 nm on the position of single grains was achieved in both $(x,y)$ coordinates. This technique was extended to a 3D reconstruction \cite{Alexandrov_patent}.

The \newsdm\ detector will be placed in the Gran Sasso Underground Laboratory, surrounded by a shield to reduce external background sources.
It will be placed on an equatorial telescope in order to keep its orientation towards the Cygnus constellation fixed, where the WIMP wind 
is supposed to come from. The emulsion films will be placed parallel to the Galactic plane.
This peculiar orientation is ideally suited for the detection of neutrinos produced by supernova explosions, since these phenomena are expected to occur in the galactic plane, where the star density is higher.

The emulsion reference frame used in
this paper has the $X_e$ and $Y_e$ axes pointing towards the Cygnus constellation and the Galactic Centre, respectively.
The $Z_e$ axis is perpendicular to the Galactic Plane.
Spherical coordinates $(\phi_e,\theta_e)$ have been used to indicate the direction of the induced nuclear recoils: 
$\theta_e$ is the inclination angle from the galactic
$X_e$-$Y_e$ plane 
while $\phi_e$ is the angle in the galactic plane from the Cygnus direction
\begin{equation}
    \theta_e \in \left[-\frac{\pi}{2},\frac{\pi}{2}\right]
    \qquad\phi_e\in \left(-\pi,\pi\right].
\end{equation}
Details about the reference system, the coordinate systems and their relationships are described in Appendix \ref{app:APPENDIX}.

\section{Neutrinos and background sources}\label{sec:sources}

Simulation studies based on \geant\ toolkit \cite{Agostinelli:2002hh} have been performed to  predict the interaction rate in the detector due to neutrinos emitted by a supernova and to estimate the track length and the angular distributions of the induced recoils. On the other hand, the supernova neutrino signal can be mimicked mainly by two background sources:
radiogenic neutrons and solar neutrinos from \textsuperscript{8}B. 
Neutron induced recoils  cannot be distinguished from the CE$\nu$NS in the detector.  Moreover, 
ton scale detectors for dark matter search are also sensitive to solar neutrinos.

\subsection{Supernova neutrino signal}
Neutrinos coming from a supernova explosion can produce tracks in NIT emulsions with lengths depending on the kinetic energy transferred 
to the recoiled nucleus (see Eq.\ \ref{eq:Emin}). The number of expected events depends on the detector threshold, the exposed mass and the distance of the source 
from the Earth.

\begin{figure}[t]
\centering
\begin{subfigure}[b]{0.48\textwidth}
\includegraphics[width=\textwidth]{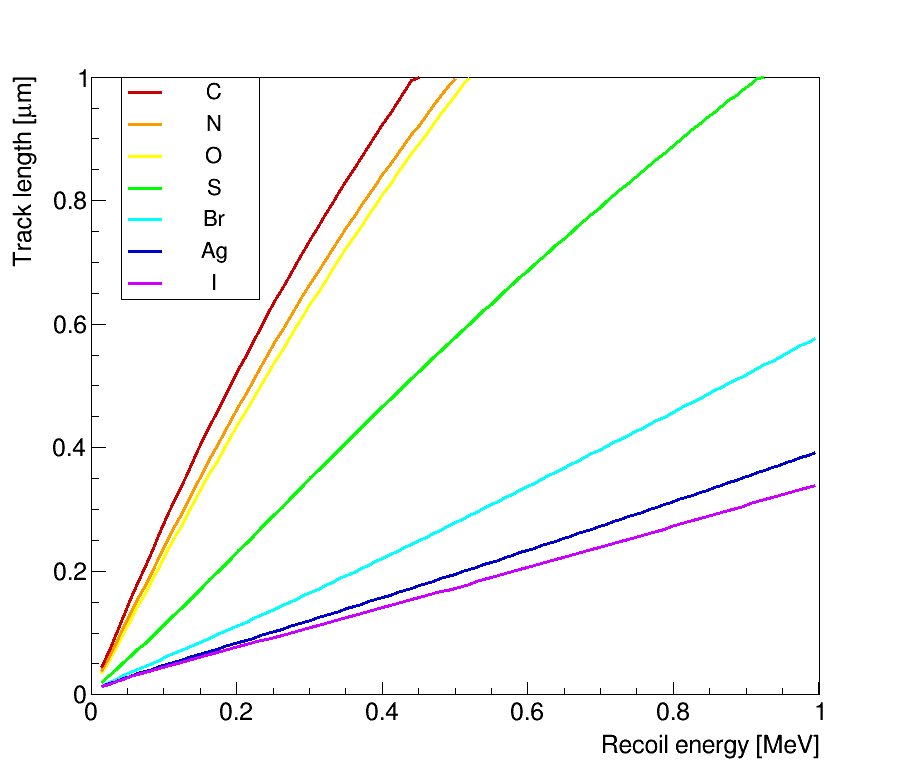}
\caption{Track length}
\label{fig:ene_len}
\end{subfigure}
\begin{subfigure}[b]{0.48\textwidth}
\includegraphics[width=\textwidth]{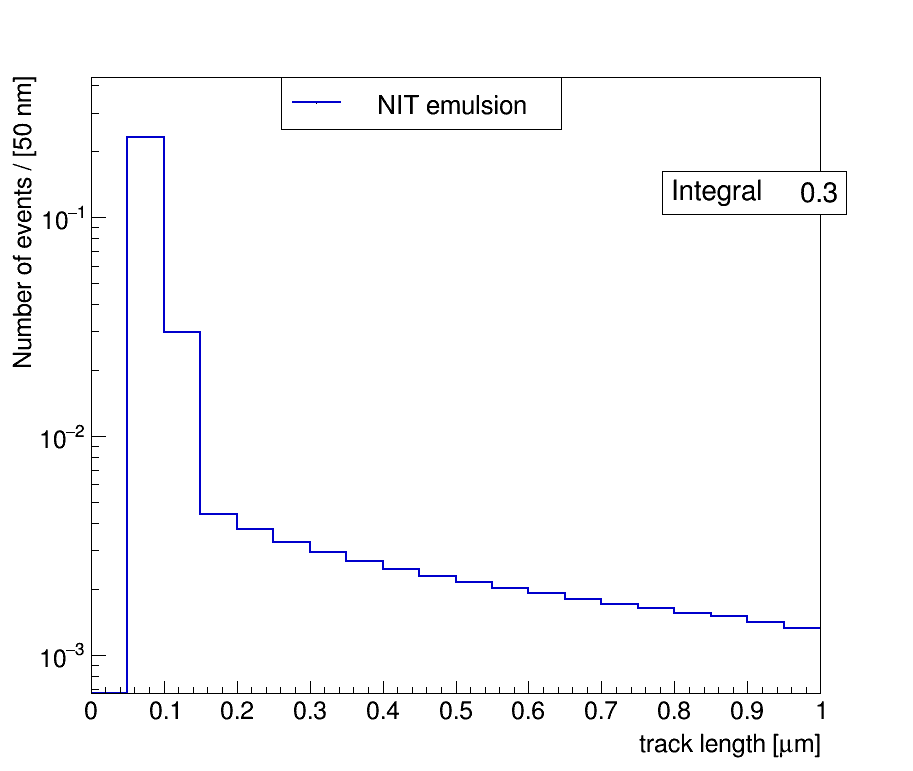}
\caption{Number of events}
\label{fig:trklen}
\end{subfigure}
\caption{
    (\protect\subref{fig:ene_len})
    Track length versus transferred energy
	for the target nuclei of NIT emulsions.
	Lighter nuclei can produce track lengths up to a few
	microns, while heavier ones only up to a few hundreds
	nanometers. (\protect\subref{fig:trklen}) Track
	length distribution in NIT emulsions of supernova neutrino induced recoils 
	in the range $[0.05,1]$ \si{\micro\meter}.
	The distribution is normalised to the number of 
	expected events in one ton of active mass.
}
\label{fig:eve:tot}%
\end{figure}

For nuclear emulsion detectors the energy threshold corresponds to the minimum detectable track length. The correlation between the transferred energy to 
the recoiled nucleus and the expected range in emulsions has been evaluated for each target element using SRIM \cite{2010SRIM}.
The track length $L$ of nuclear recoils induced by supernova neutrino scattering versus the transferred energy is reported in Fig.\ \ref{fig:ene_len} for each target nucleus.
Only energies corresponding to recoil track lengths equal or larger than \SI{50}{\nano\meter} have been considered for the integral in Eq.\ \ref{eq:dNdK}.
Assuming the parameters defined in Sec.\ \ref{sec:model}, the total number of expected events in the whole detector is
\SI{0.30}{\per\ton}.

Distributions of CE$\nu$NS kinematic variables have been simulated through a Monte Carlo generation.
The track length distribution of the induced recoils normalized to the number of expected events is reported in Fig.\ \ref{fig:trklen} in the $[0.05,1]$ \si{\micro\meter}
range.

The scattering angle $\theta_{sc}$
is defined as the angle between the incoming neutrino
direction and the one of the scattered nucleus. Its value
can be obtained by the following relation:
\begin{equation}
 \cos\theta_{sc} =  \frac{E_{\nu} + M}{E_{\nu}}\sqrt{\frac{K}{2M}},
\end{equation}
where $K$ is the transferred kinetic energy in the non-relativistic limit ($|\vec{v}_{rec}| = \sqrt{2K/M}$).

Assuming a supernova explosion from the direction of the 
Galactic Centre, the scattering angles of neutrino induced recoils are generated and the angular distributions of $\theta_e$ and  $\phi_e$ derived as 
reported in Fig.\ \ref{fig:angTheta} and \ref{fig:angPhi}
(see Appendix \ref{app:APPENDIX}). 

\begin{figure}[t]
\centering
\begin{subfigure}{0.45\textwidth}
\includegraphics[width=\textwidth]{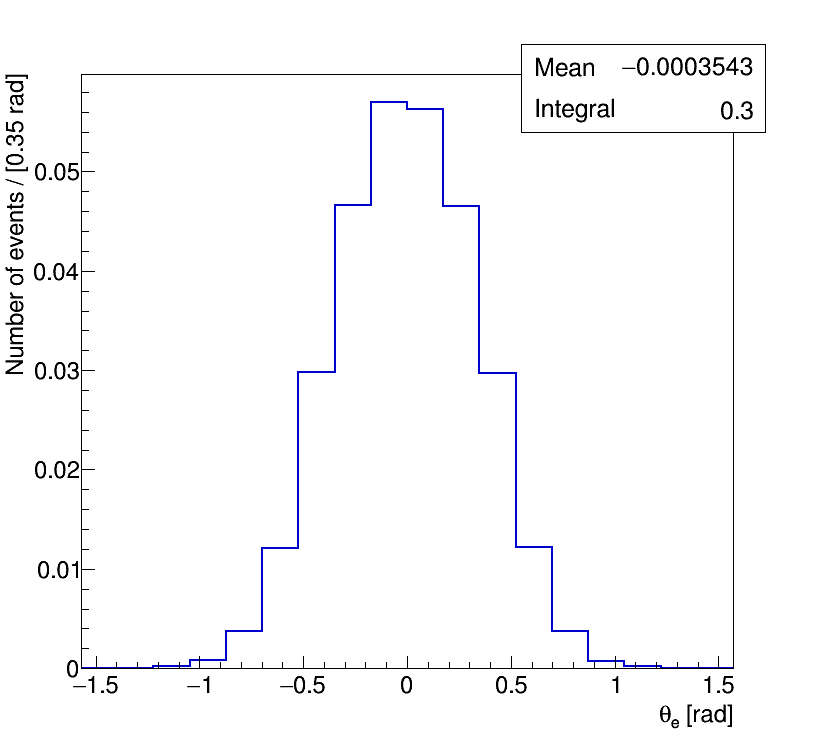}
\caption{$\theta_{e}$ distribution}
\label{fig:angTheta}
\end{subfigure}
\hfill
\begin{subfigure}{0.45\textwidth}
\includegraphics[width=\textwidth]{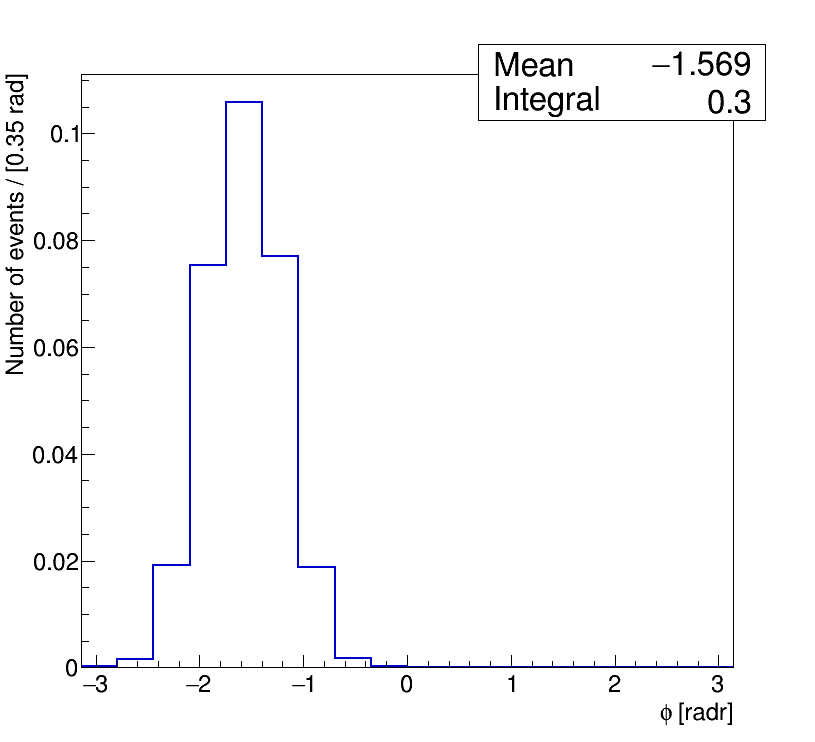}
\caption{$\phi_{e}$ distribution}
\label{fig:angPhi}
\end{subfigure}
\caption{
Distributions of the emulsion angles
$\theta_e$ (\subref{fig:angTheta}) and $\phi_e$
(\subref{fig:angPhi}) of supernova
neutrino induced recoils in the \newsdm\ detector.
The induced recoils are scattered mainly in the direction ($\theta_e=0$) and opposite to the
incoming supernova neutrino direction ($\phi_e=-\pi/2$).}
\label{fig:ang}%
\end{figure}

In this study it is assumed that the incoming neutrino direction points to the Galactic Centre, since it is the region where the stellar mass density is higher and therefore there is a higher probability for a supernova explosion; note that this argument is supported by the distribution of the historical records of supernova remnants and supernovas, see e.g.\ Ref.\ \cite{snr}.

The $\theta_e$ angle distribution is peaked at zero since induced recoils are mostly diffused around the neutrino incoming direction; the $\phi_e$ angle distribution is peaked at $-\pi/2$ rad, along the negative $-Y_e$ axis corresponding to the direction opposite to the Galactic Centre.

\subsection{Background sources}

Background from $\beta$-rays produced in \textsuperscript{14}C decays is expected to be negligible  since NIT emulsions are hardly sensitive to electron recoils, in particular when cooled down to $\sim\SI{100}{\kelvin}$ \cite{Kimura:2017hme}. The replacement of gelatin with synthetic polymers is also being studied.
The main background sources is therefore induced by neutrons.

Neutron-induced recoils from intrinsic contamination in NIT emulsions were recently evaluated with a \geant\ simulation, giving a yield of $\sim 1$ neutron \si{\per\kilo\gram\per\year} \cite{Aleksandrov:2015pfa}. 
This study has also shown that the neutron-induced events can be reduced down to \SI{0.06}{\per\kilo\gram\per\year} exploiting directionality
and track length cuts. This result was obtained without any purification of the materials involved \cite{Aleksandrov:2015pfa}.

Nowadays ton-scale detectors for direct dark matter search have made  the external neutron background negligible using appropriate shields. For what concerns intrinsic background 
materials are used for ultra-low radioactivity detectors. 
As a result of this effort the radiogenic neutrons are typically reduced down to the level of $\sim 1$ 
neutron \si{\per\ton\per\year}. In this study we assume that 
\newsdm\ will reach the same high-purity standards.

The track lengths of supernova neutrino induced recoils are at least a few hundred nanometers long. 
For this reason, only track lengths shorter than \SI{1}{\micro\meter} can be dangerous for the signal observation. 
The number of neutron-induced recoils with track lengths in the $[0.05,1]$ \si{\micro\meter}  range amounts to
$\sim \SI{0.33}{\per\ton\per\year}$, where fiducial
volume effects have been accounted for.

\subsection{\texorpdfstring{Solar neutrinos from \textsuperscript{8}B}{Solar neutrinos from 8B}}
\begin{figure}[!t]
\centering
\includegraphics[width=0.7\textwidth]{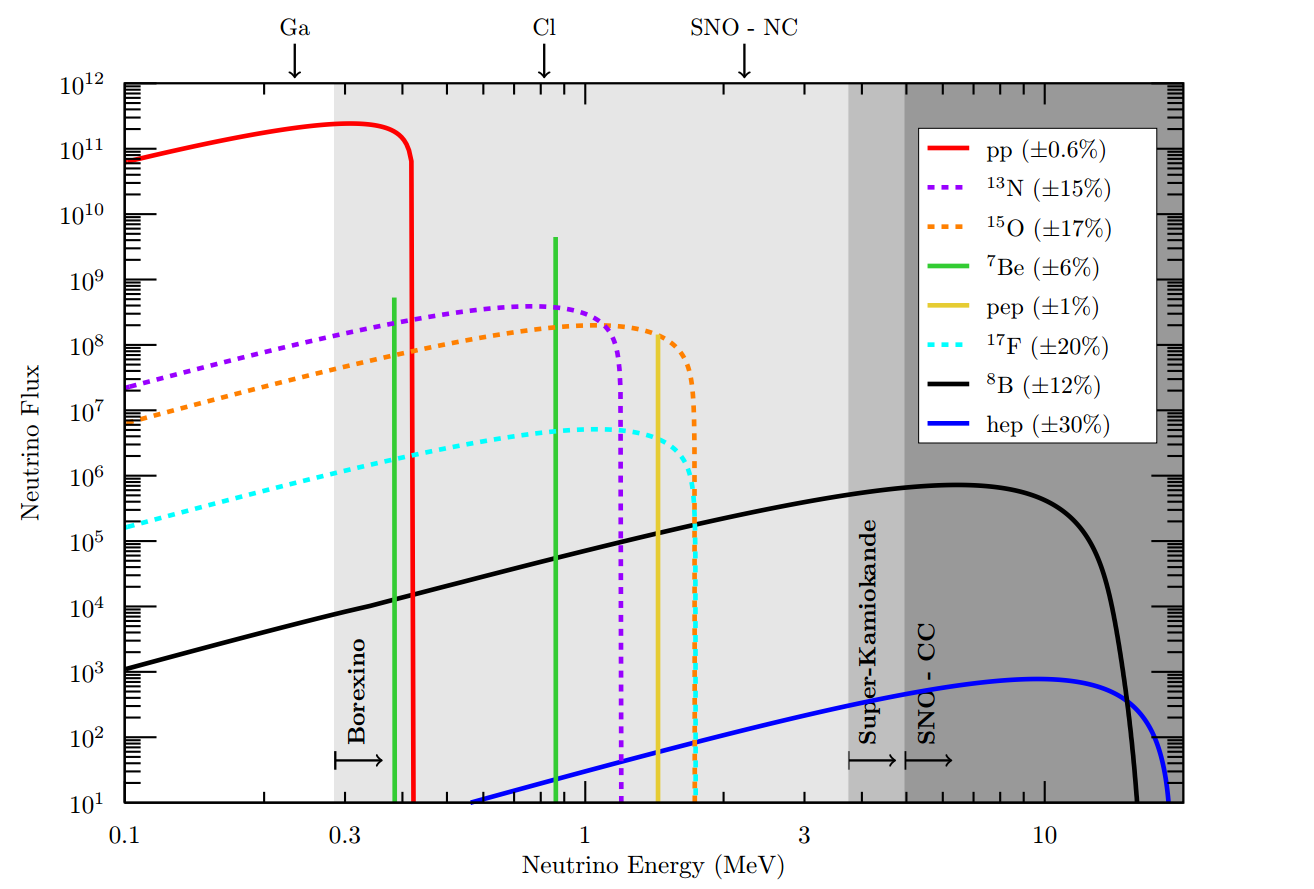}
\caption{\label{fig:Bflux} Solar neutrino flux
$\df\Phi_{\nu}/\df E_{\nu}$ for each source as a function of
the energy \protect\cite{ssm} along with the sensitivity regions of the various solar neutrino experiments \cite{solarflux}.} 
\includegraphics[width=0.7\textwidth]{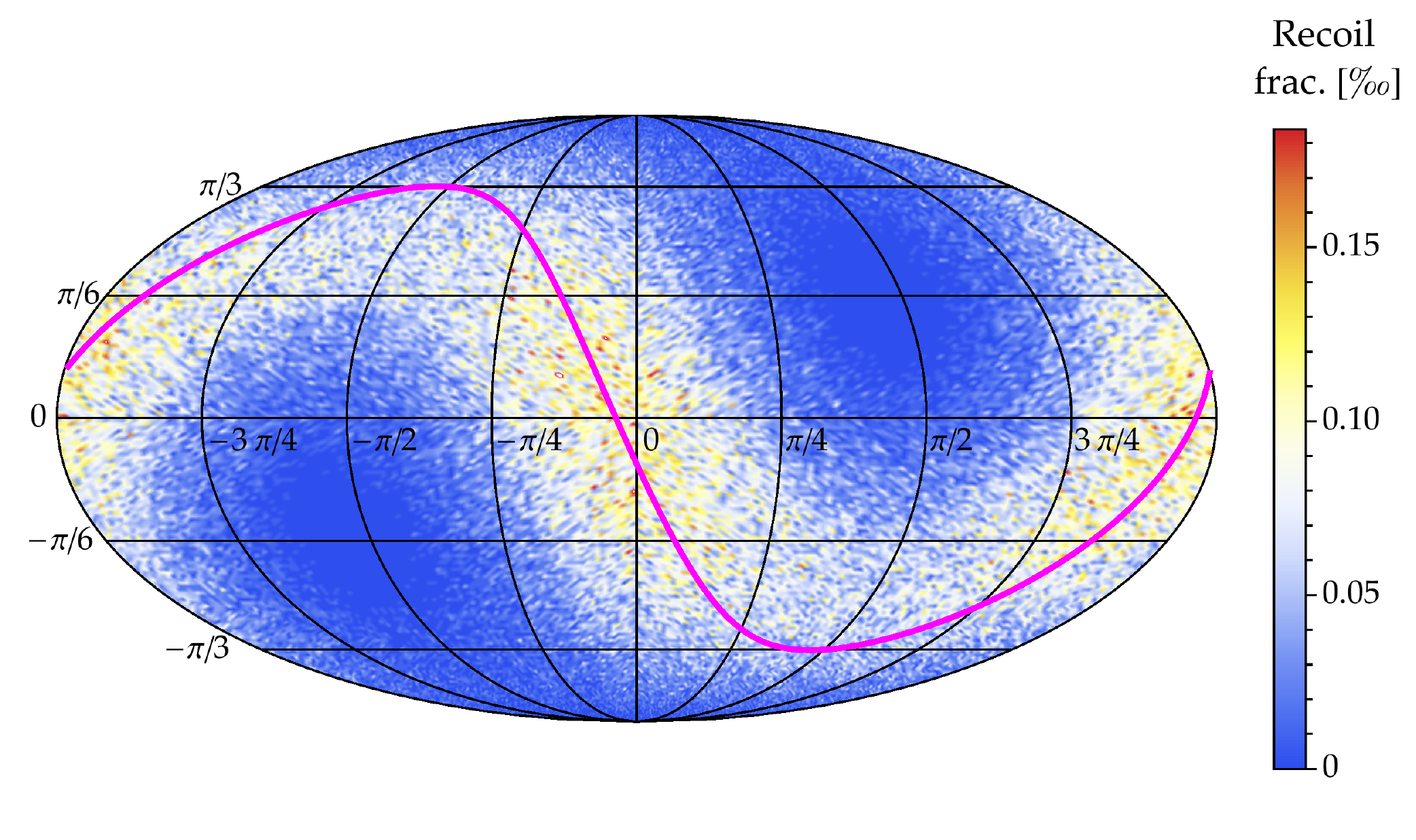}
\caption{Mollweide projection in a Galactic-like coordinate
system of the induced recoils from \textsuperscript{8}B solar
neutrinos: the latitude corresponds to $\theta_e$ while
the longitude to $\phi_e - \pi/2$
(see Appendix~\protect\ref{app:APPENDIX}).
The magenta line marks the neutrino arrival direction,
i.e.\ from the Sun to the Earth.}
\label{fig:mollweide}
\end{figure}

Ton-scale mass detectors are sensitive also to the recoils
induced by solar neutrinos from \textsuperscript{8}B
whose energies extend up to $\sim\SI{16}{\mega\electronvolt}$
as shown in Fig.\ \ref{fig:Bflux}.
The total rate of expected events is given by
\begin{equation}
\frac{\df{\mathrm{N}}}{\df{t}} = N_T \iint_{E_\mathrm{inf}} \df{\Omega} \df{E_{\nu}} \frac{\df{\sigma}}{\df{\Omega}}\frac{\df{\Phi_{\nu}}}{\df{E_\nu}} 
\end{equation}
where $\df\Phi_{\nu}/\df E_{\nu}$ is the differential solar neutrino flux (see Fig. \ref{fig:Bflux}). 

Note that the \textsuperscript{8}B solar neutrino flux has been measured by SNO with neutral currents \cite{sno} and the result  $\Phi^{\text{\textsc{sno}}}_{\text{\textsc{b}}}=
\SI[separate-uncertainty=true]{5.25 +- 0.21e6}{\per\centi\meter\squared\per\second}$ is more precise than the
theoretical predictions and it is independent upon neutrino oscillations.

The direction of incoming neutrinos from \textsuperscript{8}B has been simulated  using the projection of the Earth velocity around the Sun onto galactic axes 
\cite{1475-7516-2014-02-027}. 
The Earth revolution orbit is assumed to be circular. 
The Mollweide projection in a Galactic-like coordinate system of the induced recoils, assuming one ton per year exposure, is reported in Fig.~\ref{fig:mollweide} where the magenta line represents the revolution of the Earth around the Sun.
The number of expected induced recoils with track lengths in the $[0.05,1]$ \si{\micro\meter} range 
amounts to $\sim\SI{0.18}{\per\ton\per\year}$. The observation of \textsuperscript{8}B neutrinos would be relevant as a control sample.

\section{Results}

\begin{figure}[t]
\centering
\includegraphics[width=8cm]{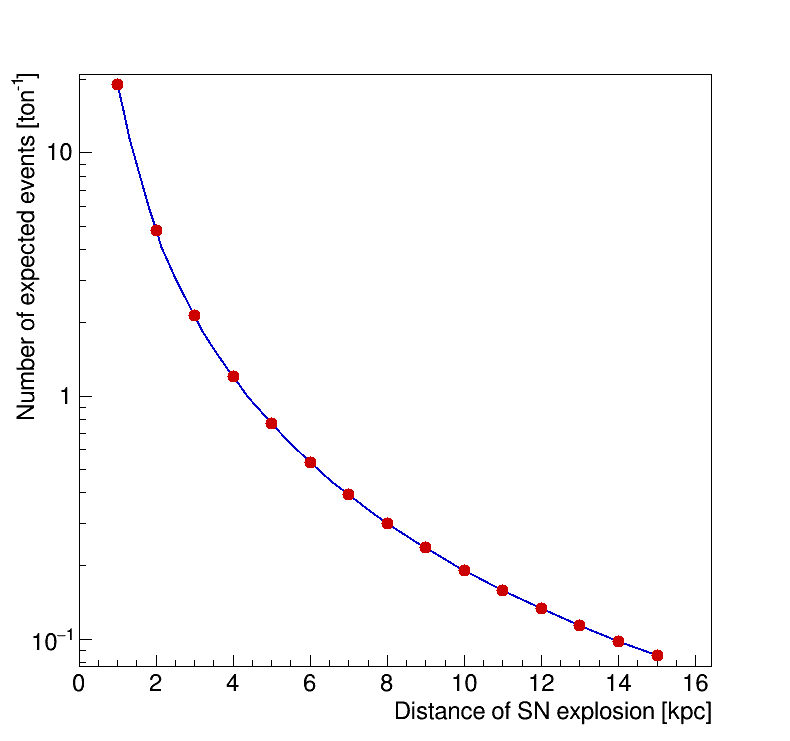}
\caption{
    Number of expected events per ton of active mass
    as a function of the
    distance $D$ of the supernova explosion. 
}
\label{fig:ev_dist}
\end{figure}

The expected number of nuclear recoils induced by neutrinos and neutrons is proportional
to the target mass. Being the supernova explosion a transient phenomenon, the signal from supernova neutrinos does not depend on the exposure time but rather on the inverse of the squared distance $D$, as shown in Fig.\  \ref{fig:ev_dist} assuming a \SI{50}{\nano\meter} threshold in the detection of nuclear recoil tracks. 
On the contrary, the background from solar neutrinos and neutrons is proportional to the exposure time.

The total number of expected events depends also on the detection threshold.
The potentiality of the observation of the supernova neutrinos was studied assuming a signal region ranging from \SI{50}{\nano\meter} to \SI{1}{\micro\meter}, for a detector mass of \SI{30}{\ton} and a distance $D$ of \SI{8}{\kilo\parsec}. The corresponding number of expected events is:
\begin{equation}
    \mathrm{N}\left(\nu\middle|\mathrm{SN}\right) = 9.0.
    \label{eq:atteseSN}
\end{equation}
On the other hand, the rate due to
\textsuperscript{8}B neutrinos and
background neutrons \Pneutron due to intrinsic contamination (IC) is:
\begin{align}
    \label{eq:bkg1}
    \mathrm{N}(\nu,\text{\textsuperscript{8}B})
    &= \SI{5.4}{\per\year};\\
    \mathrm{N}\left(\Pneutron, \mathrm{IC}\right)
    &= \SI{9.9}{\per\year}.
    \label{eq:bkg2}
\end{align}

\subsection{Supernova neutrinos observation}\label{sec:sn_results}
In order to evaluate the capability of the \newsdm\ detector to distinguish supernova neutrinos from background events,
a Profile Likelihood \cite{Cowan:2010js} ratio test has been performed. The null hypothesis $H_{0}$ (background only) has been tested against the alternative
hypothesis $H_{1}$ (signal plus background). The extended likelihood function can be written as:
\begin{equation}\label{eq:lik}
 \mathcal{L} = \frac{(\mu_s + \mu_b)^N}{N!}e^{-(\mu_s + \mu_b)} \times \prod_{i=1}^N \left(\frac{\mu_s}{\mu_s + \mu_b}
 \prod_j S(x_{ij}) + \frac{\mu_b}{\mu_s + \mu_b} \prod_j B(x_{ij}) \right),
\end{equation}
where $n$ is the total number of
observed events, $\mu_s$ is the number of expected supernova neutrinos given in Eq.\ \eqref{eq:atteseSN} and $\mu_b$
the expected background, obtained by multiplying the
expected rates \eqref{eq:bkg1} and \eqref{eq:bkg2} by
the the exposure time $\Delta T$ (see e.g.\ Fig.\ \ref{fig:significanze}):
\begin{equation}
    \mu_b = \left[\mathrm{N}(\nu,
    \text{\textsuperscript{8}B}) +
    \mathrm{N}\left(\Pneutron, \mathrm{IC}\right)\right]
    \Delta T.
\end{equation}
For the $i$-th event, the value $x_{ij}$ is the $j$-th variable used in the test statistics while
the functions $S(x)$ and $B(x)$ are the probability density functions (PDF) for signal and background, respectively.

\begin{figure}[t]
\centering
\begin{subfigure}{0.3266\textwidth}
\includegraphics[width=\textwidth]{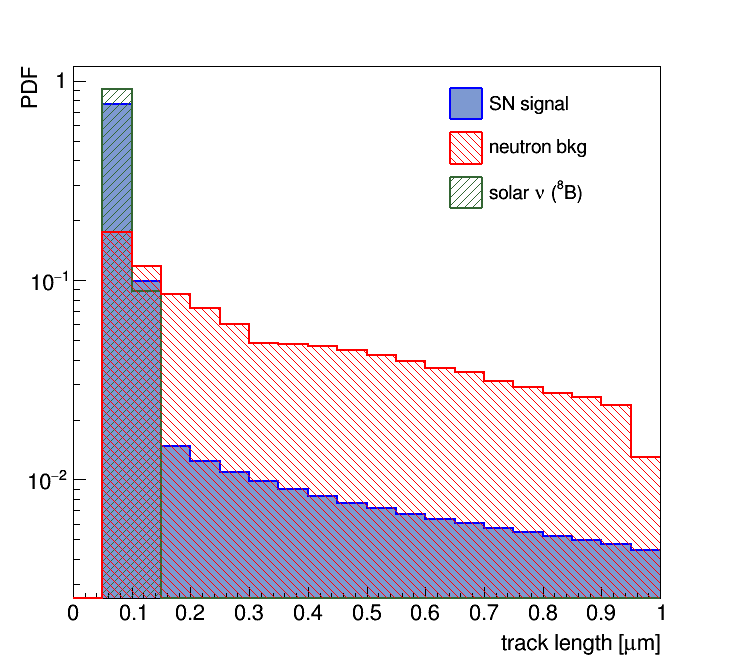}
\caption{$L$}
\label{fig:rlung}
\end{subfigure}
\hfill
\begin{subfigure}{0.3266\textwidth}
\includegraphics[width=\textwidth]{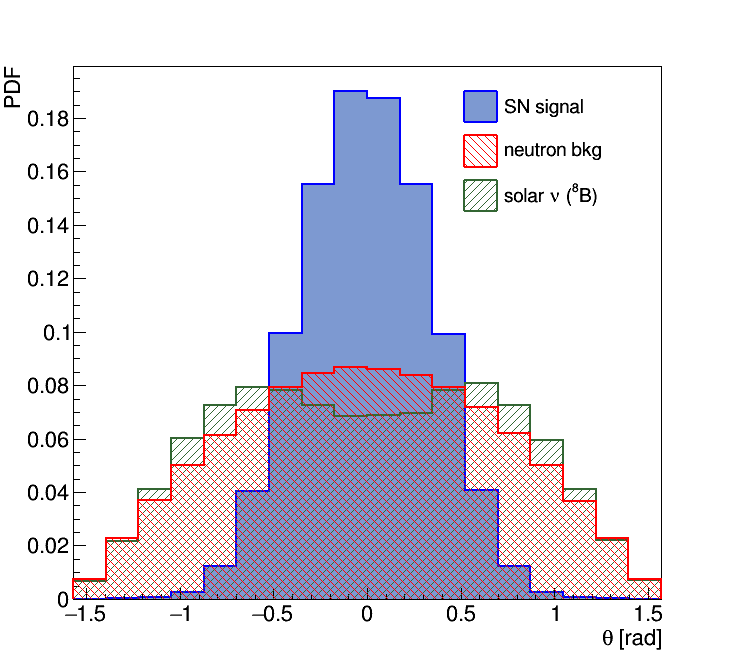}
\caption{$\theta_e$}
\label{fig:rtheta}
\end{subfigure}
\begin{subfigure}{0.3266\textwidth}
\includegraphics[width=\textwidth]{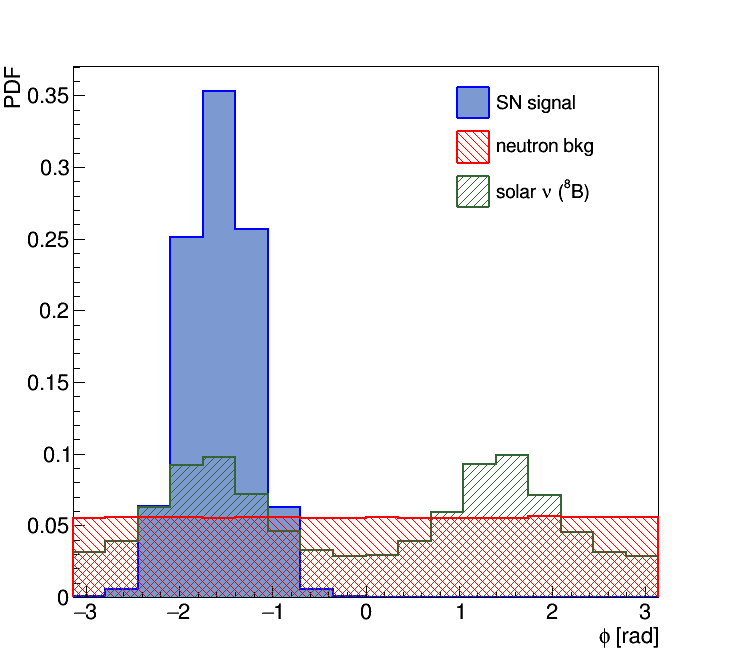}
\caption{$\phi_e$}
\label{fig:rphi}
\end{subfigure}
\caption{
\label{fig:rpdf}%
PDFs of recoiled nuclei, normalized to 1, induced by supernova neutrinos (blue), solar neutrinos from \textsuperscript{8}B (green)
and radiogenic neutrons
(red). The track length, $\theta_e$ and $\phi_e$ distributions are shown in panels (\protect\subref{fig:rlung}), (\protect\subref{fig:rtheta}) and (\protect\subref{fig:rphi}), respectively.}
\end{figure}

\begin{figure}[t]
\centering
\begin{subfigure}{0.45\textwidth}
\includegraphics[width=\textwidth]{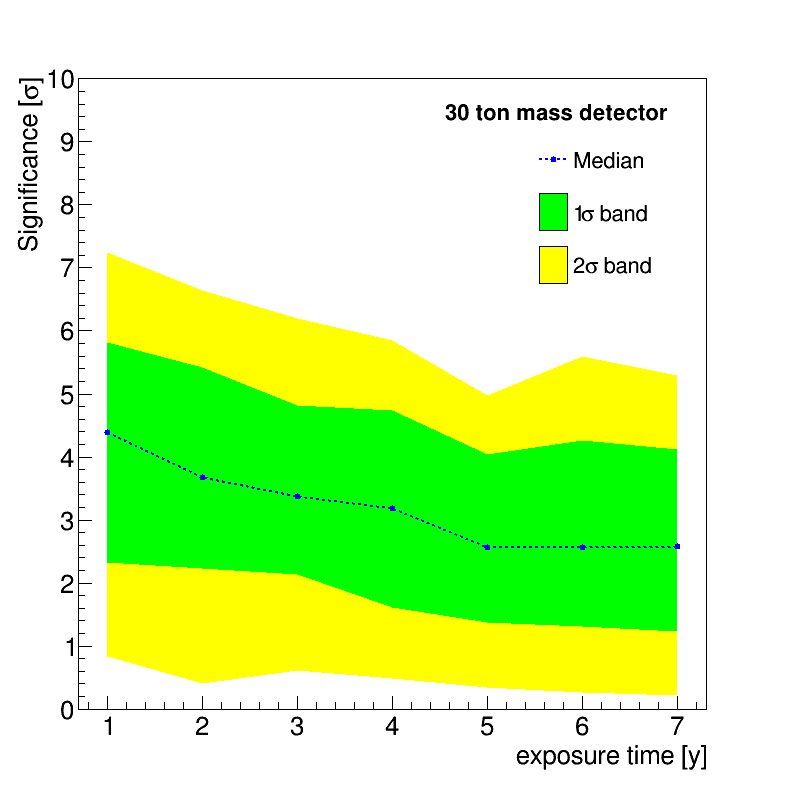}
\caption{SN signal}
\label{fig:signif}
\end{subfigure}
\hfill
\begin{subfigure}{0.45\textwidth}
\includegraphics[width=\textwidth]{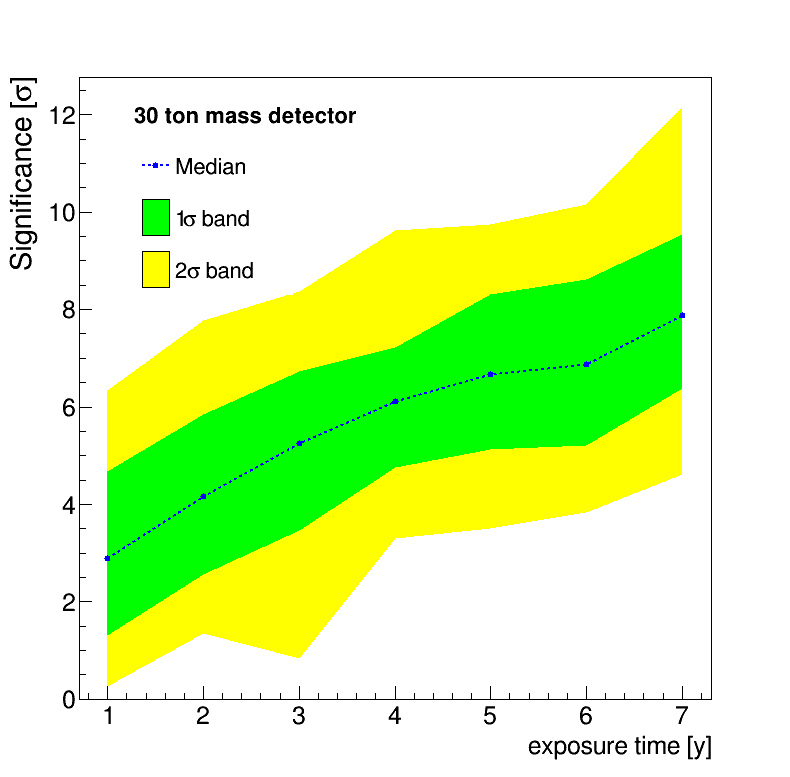}
\caption{\textsuperscript{8}B neutrinos}
\label{fig:boro_test}
\end{subfigure}
\caption{
Mean significance as a function of the exposure time for a
\SI{30}{\ton} mass detector (blue dotted line),
for the supernova neutrino signal
(\protect\subref{fig:signif}) and the \textsuperscript{8}B
neutrinos (\protect\subref{fig:boro_test})
The green and yellow solid colour regions show the 
confidence levels at 68\% and 95\% respectively.
\label{fig:significanze}%
}
\end{figure}

A set of three variables has been used: the track length $L$, the $\phi_e$ and $\theta_e$ angles of the induced recoils.
The PDFs of the three variables for supernova neutrinos, solar neutrinos from \textsuperscript{8}B
and intrinsic neutrons are reported in Fig.\ \ref{fig:rpdf}.

The significance of the test statistics for the signal plus
background hypothesis (S+B) has been studied using the
ROOFIT toolkit \cite{Verkerke:2003ir}.
All the three above mentioned variables have been used
in the Profile Likelihood function.
Fig.\ \ref{fig:signif}  shows the mean
significance as a function of the exposure time assuming
30 ton detector mass and  a distance $D$ of
\SI{8}{\kilo\parsec}.
The median expectation for the supernova neutrino signal is
represented by the blue dotted line with the green (68\% CL)
and yellow (95\% CL) solid colour regions. 
The shorter the exposure time the larger the significance of
S+B hypothesis, since the background increases with the time while the supernova explosion is a transient event. 
An observation of supernova neutrinos with a confidence level larger than 
$3\sigma$ requires that the detector has been operated --- i.e., that background events has been collected for a time shorter than 4 years.

\subsection{Supernova characterisation}

\begin{figure}[t]
\centering
\includegraphics[width=8cm]{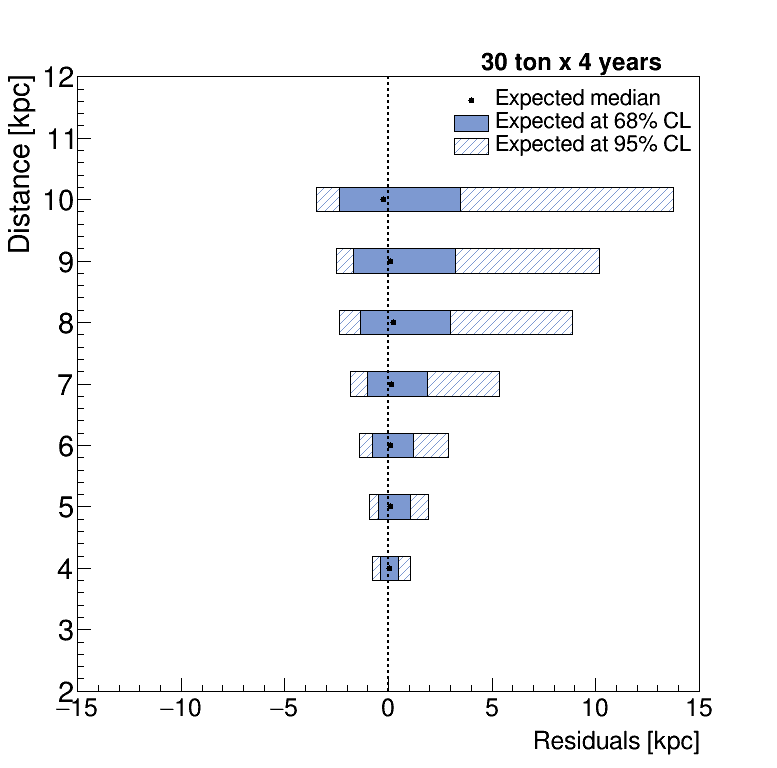}
\caption{Residuals between the measured and expected distance.}
\label{fig:res}%
\end{figure}

In addition to the signal observation, the directionality can be exploited to retrieve important information on the supernova.
The likelihood function in Eq.\ \ref{eq:lik} can indeed be used to derive the distance of the supernova explosion which is proportional to the inverse square root of the number of observed signal events.\footnote{We assume the correctness
of the model exposed in Sec.\ \ref{sec:model}. Note that the hypothesis on the model can be validated using the combination of different detection channels, in particular those by Super-Kamiokande and Hyper-Kamiokande \cite{GalloRosso:2017mdz, GalloRosso:2017hbp}.}
One thousand of pseudo-experiments were simulated for different distances and a fit of maximum likelihood was used to extract the  number of signal events ($\mu_s$) and therefore the measured distance $D$ after two years of detector operation. 
The residual between the measured and expected distance is reported in Fig.\ \ref{fig:res} with 68\% and 95\% C.L. intervals. The expected median is centred at zero and the measurement is more precise for shorter distances where the signal is expected to be larger.

\subsection{\texorpdfstring{Solar neutrinos from \textsuperscript{8}B as a control sample}{Solar neutrinos from 8B as a control sample}}

The capability of the \newsdm\ detector to identify
nuclear recoils induced by solar neutrinos from
\textsuperscript{8}B 
has been also studied. Fig.\ \ref{fig:boro_test} shows the
mean significance as a function of the exposure for \SI{30}{\ton} mass detector.
The median expectation for solar neutrinos from
\textsuperscript{8}B signal is represented
by the blue dotted
line with the green (68\% CL) and yellow (95\% CL) solid 
colour regions. The longer the exposure time the larger the
significance of S+B hypothesis, since the signal-to-background ratio 
increases with time.

After two years, an observation of the signal from \textsuperscript{8}B solar neutrinos can be achieved with a confidence level of 3$\sigma$ .

\section*{Conclusions}\label{sec:con}
The next generation of ton-scale detectors for dark matter searches will be sensitive to different neutrino sources like solar neutrinos from boron and neutrinos coming from a supernova explosion. Coherent elastic neutrino-nucleus scattering events are very intriguing to observe for a dark matter detector, however they can also represent a serious source of background events, unless directional information can be exploited. 
The capability of identifying neutral current interactions of neutrinos originated by a supernova explosion has been studied using a directional detector based on nuclear emulsions with nanometric crystals. 
The \newsdm\ experiment uses emulsion films produced at Nagoya University made of silver halide crystals with nanometric size which make it possible to reconstruct track lengths down to \SI{50}{\nano\meter} when readout with appropriate optical microscopes.
Assuming the emission parameters from numerical simulations, the expected number of supernova neutrino induced recoils with path lengths ranging 
from \SI{50}{\nano\meter} to \SI{1}{\micro\meter} is about \SI{0.30}{\per\ton} for a nominal supernova signal as defined in Sec.\ \ref{sec:model}. The track length and the angular distributions have been predicted through a toy Monte Carlo and a Likelihood Ratio test has been performed to discriminate the signal from the background sources represented by the intrinsic neutrons and solar neutrinos from boron. 
The observation of neutrinos from a supernova exploded at
\SI{8}{\kilo\parsec} can be obtained with a \SI{30}{\ton} detector only in the first few years of operation:
as an example, if the supernova will explode within four years, a $3\sigma$ signal can be achieved.
In this scenario, the distance of the supernova explosion from the Earth can be determined and the expected precision increases strongly for smaller distances.
A Likelihood Ratio test has been also performed to discriminate recoils induced by \textsuperscript{8}B solar neutrinos from neutron induced recoils for a control sample. A significance larger than 3$\sigma$ CL is obtained with an exposure larger than two years for a detector of \SI{30}{\ton}. 

\appendix
\section{Coordinate transformations}
\label{app:APPENDIX}

The calculation described in this paper
make necessary coping with different coordinate
systems, especially when considering the
modulation rate of solar neutrinos. 
In order to make this tricky
point as clear as possible, the relation between
these systems are described in the following.

\subsection{Coordinate systems}
\label{app:coord}
\begin{figure}[t]
\centering
\begin{subfigure}[b]{0.39\textwidth}
\includegraphics[width=\textwidth]{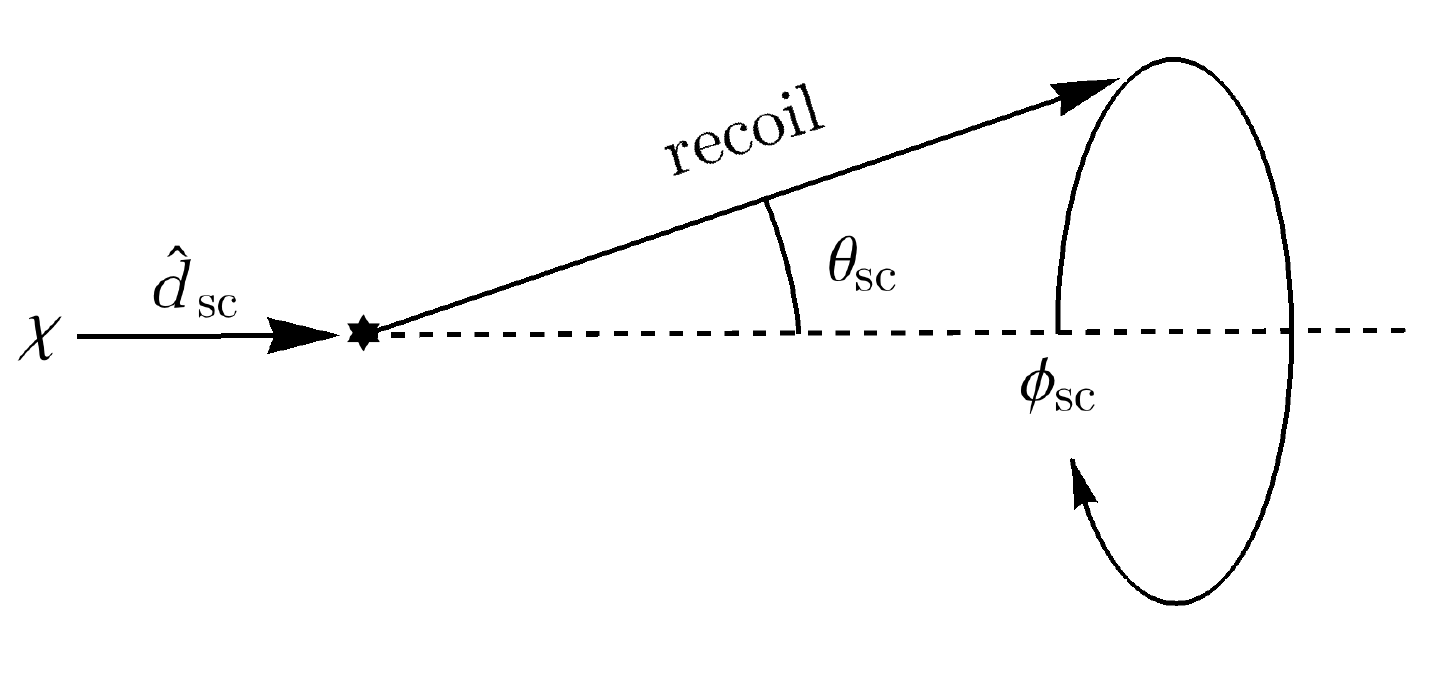}
\caption{Scattering R.F.}
\label{fig:RFsca}
\end{subfigure}
\begin{subfigure}[b]{0.29\textwidth}
\includegraphics[width=\textwidth]{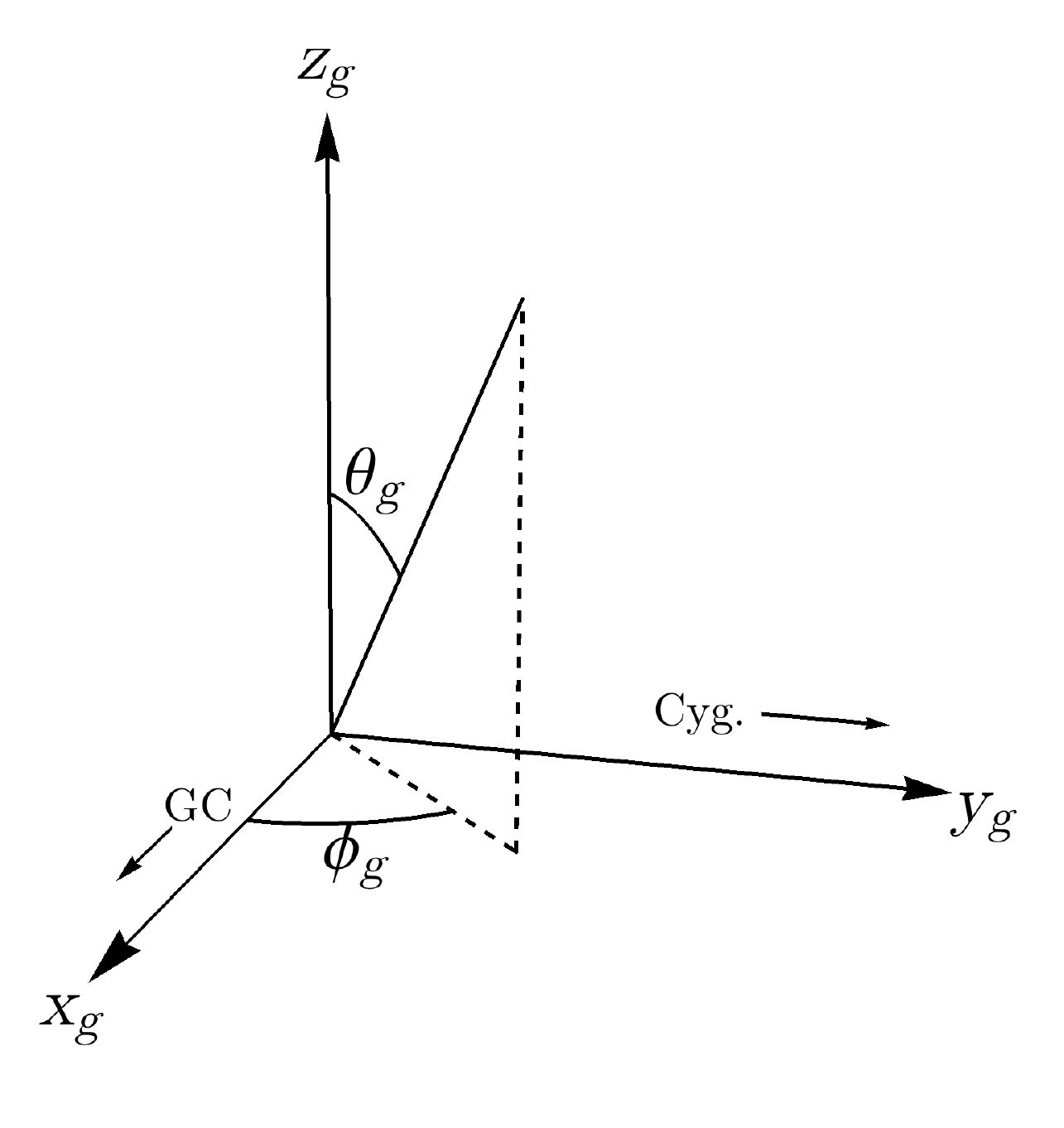}
\caption{Standard Galactic R.F.}
\label{fig:RFgal}
\end{subfigure}
\begin{subfigure}[b]{0.29\textwidth}
\includegraphics[width=\textwidth]{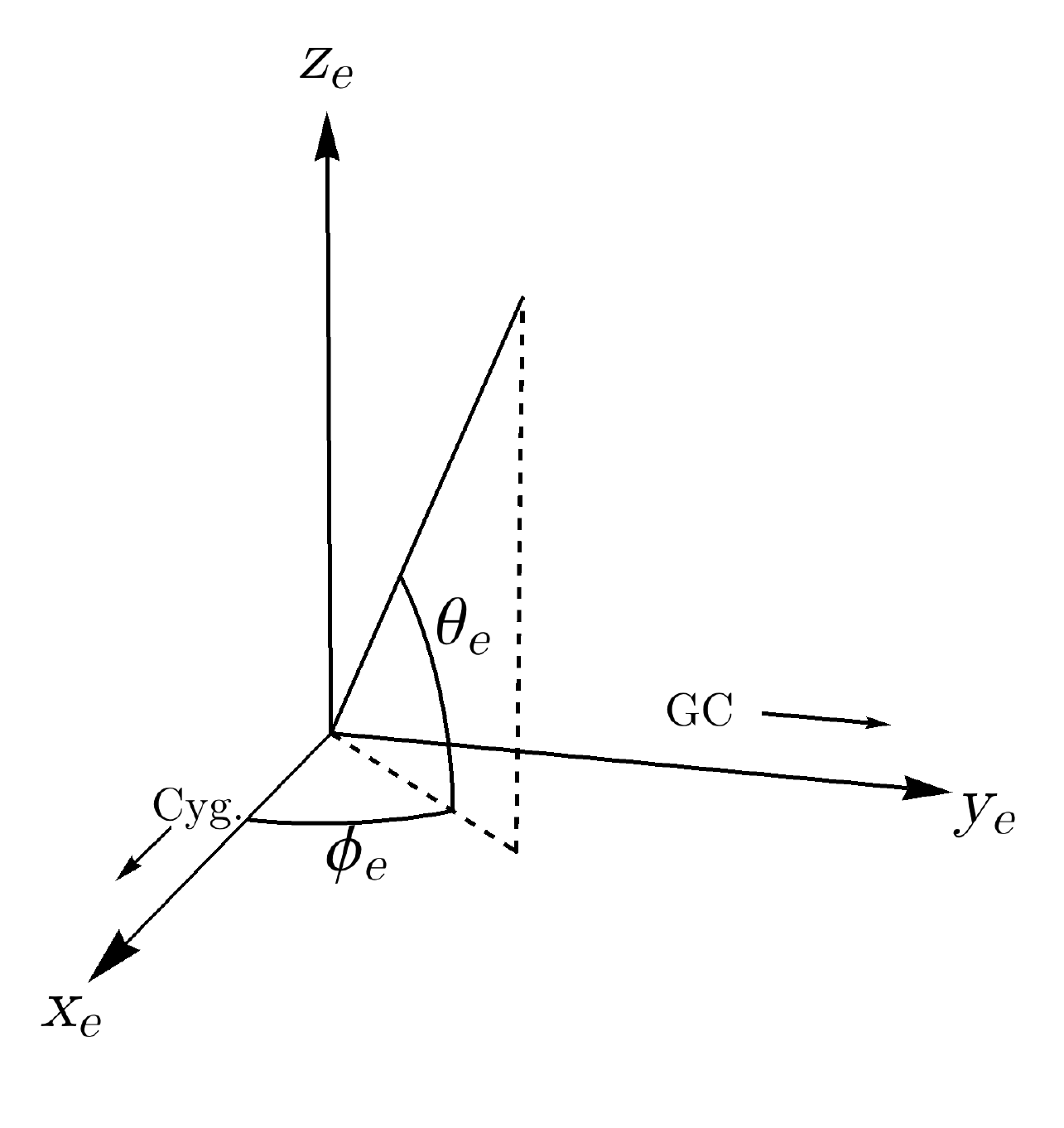}
\caption{Emulsion R.F.}
\label{fig:RFemu}
\end{subfigure}
\caption{Reference frames used in this paper. 
The interaction reference frame is described in 
(\subref{fig:RFsca}). In this frame, the particle $\chi$
--- travelling in the $\hat{d}_{sc}$ direction ---
hits a nucleus in the origin, marked by the star,
generating a "recoil cone" described by the angles
$\left(\theta_{sc}, \phi_{sc}\right)$.
The systems (\subref{fig:RFgal}) and
(\subref{fig:RFemu}) are both galactic reference frames
with the sun in the origin. 
}
\label{fig:sistemi}%
\end{figure}

In this paper they have been adopted three main (polar) reference
frames reported graphically in Fig.\ \ref{fig:sistemi}.
\begin{description}
	\item[Scattering frame:] this is the reference
	system where the interaction between the
	incident particle and the target is described
	--- see Fig.~\ref{fig:RFsca}. The origin is
	placed in the interaction point.
	A unit vector in this system is identified
	by the scattering angles $\left(\theta_{sc},
	\phi_{sc}\right)$ where $\theta_{sc}$ is the
	angle between the recoiled nucleus and the
	direction of the incident particle, and
	varies in the range $\left[0,\pi\right]$. The angle
	$\phi_{sc}$ does not depend on the interaction
	and varies uniformly in $\left[0,2\pi\right)$.
	\item[Standard galactic frame:] this is the standard
	galactic reference system, with the Sun at the origin
	(Fig.~\ref{fig:RFgal}). The $X_g$ axis
	points toward the Galactic Centre,
	the $Y_g$ axis points in the direction of
	the solar system motion (Cygnus constellation) and
	the $Z_g$ axis is orthogonal to the galactic
	$X_g$-$Y_g$ plane in such a manner to keep
	the right-handed coordinated (North Galactic
	Pole). A unit vector in this coordinate
	system is described by the 
	polar angles 
	$\left(\theta_{g},\phi_{g}\right)$
	as
	$
		\hat{u}_g = \left(\cos\phi_g \sin\theta_g,
		\,\sin\phi_g \sin\theta_g,\,\cos\theta_g
		\right),
	$ where $\theta_g\in\left[0,\pi\right]$
	and $\phi_g\in\left[0,2\pi\right)$.
	\item[Emulsion frame:] this is the NEWS-dm
	emulsion (galactic) reference frame, with
	the Sun in the origin (Fig.~\ref{fig:RFemu}).
	The $X_e$  and $Y_e$ axes
	point toward the Cygnus constellation and 
	the Galactic Centre, respectively: the
	$X$ and $Y$ axes between the Emulsion
	and the standard system
	are swapped. As a consequence the $Z_e$
	axis points in the opposite direction of $Z_g$.
	A unit vector in this coordinate
	system is described by the emulsion angles
	$\left(\theta_{e},\phi_{e}\right)$. As a convention,
	the $\theta_{e}$ angles measures the angular distance
	from the galactic $X_e$-$Y_e$ plane and thus varies
	from $-\pi/2$ to $\pi/2$. The $\phi_e$ angle
	describes the angular distance from the $X_e$ axis
	and varies in the range $\left(-\pi,\pi\right]$.
	\end{description}

The procedure followed in this paper 
starts from the $\left(\theta_{sc},\phi_{sc}\right)$
distributions of the recoiling nuclei in the scattering frame
which are then translated into the observed
$\left(\theta_{e},\phi_{e}\right)$ distributions in the
emulsion reference frame. Therefore, a coordinate transformation from
the unit vector $\hat{u}_{sc}$ describing the
recoiling direction to the unit vector 
$\hat{u}_e$ is performed. Since the latter 
depends on the galactic direction of the source,
two different cases have been taken into account. 
The first one
concerns neutrinos coming from a
supernova explosion and is described in Sec.\ \ref{app:snu}.
The second one, discussed in Sec.\ \ref{app:nuboro}, 
regards neutrinos coming from the
sun. This case is more complex since it takes
dynamical transformations depending on the
time of the year, because of the motion of the
earth around the Sun.

\subsection{Supernova neutrinos}
\label{app:snu}

According to the galactic distribution of pre-supernova
object \cite{Costantini:2005un}, the next
supernova is expected to happen not too far from the Galactic Centre.
Since the emulsion frame has its $Y_e$ axis pointed toward
the Galactic Centre, the supernova neutrino signal would
arrive from $+Y_e$ scattering towards $-Y_e$.

The coorinate relations between the scattering reference frame 
and the emulsion one are obtained considering the recoil cone reported in Fig.~\ref{fig:RFsca} with its axis $\hat{d}_{sc}$
along the $-Y_e$ axis and its vertex in the origin, i.e.\ 
the detection point.
%
The Cartesian components of the unit vector $\hat{u}_e$ are therefore extracted
as follows:
\begin{equation}\label{eq:cartSN}
    \hat{u}_{e} = 
	\begin{cases}
		x_e &= \sin\theta_{sc}\cos\phi_{sc}\\
		y_e &= -\cos\theta_{sc}\\
		z_e &= \sin\theta_{sc}\sin\phi_{sc},
	\end{cases}
\end{equation}
where the $\phi_{sc}$ angle lies on the
$X_e$-$Z_e$ plane and starts from the $X_e$ axis.
The Cartesian coordinates in
Eq.\ \eqref{eq:cartSN} are then used 
to obtain the $\left(\theta_{e},\phi_{e}\right)$ angles:
\begin{align}\label{eq:fiSN}
	\phi_e &=
\left\{
\begin{array}{@{}l@{\quad}l@{\quad}r@{}}
	\,\tan^{-1}\left(y_e/x_e\right) &
	& \text{if $x_e > 0$}\\
	\begin{array}{l}
		\tan^{-1}\left(y_e/x_e\right) +\pi\\
		\tan^{-1}\left(y_e/x_e\right) -\pi\\
	\end{array} &
	\left.
	\begin{array}{l}
		\text{if $y_e\ge 0$}\\
		\text{if $y_e < 0$}\\
	\end{array}\right\} & \text{if $x_e < 0$}\\
	\begin{array}{l}
		\pi/2\\
		-\pi/2\\
	\end{array} &
	\left.
	\begin{array}{l}
		\text{if $y_e > 0$}\\
		\text{if $y_e < 0$}\\
	\end{array}\right\} & \text{if $x_e = 0$}\\
\end{array}
\right.\\
\theta_e &= 
\begin{cases}
\left[1 - \mathrm{sig}\left(z_e\right)\cdot 1\right]\pi/2
&\text{if $x_{e}=y_e=0$}\\
\tan^{-1}\left(z_e/
\sqrt{x_e^2+y_e^2}\right)
&\text{otherwise}\\
\end{cases}
\label{eq:tetaSN}
\end{align}
where $\mathrm{sig}{z_e}$ is the sign
of the $z_e$ coordinate.
The previous definitions preserve the range of
$\theta_e\in\left[-\pi/2,\pi/2\right]$ and
$\phi_e\in\left(-\pi,\pi\right]$.\footnote{It requires that
$\phi_e$ is in principle undefined for $x_e = y_e =0$.}
Inserting the explicit
expression of the coordinates \eqref{eq:cartSN}
in \eqref{eq:fiSN} and \eqref{eq:tetaSN}
the coordinate functions
$\theta_e\left(\theta_{sc},\phi_{sc}\right)$ and
$\phi_e\left(\theta_{sc},\phi_{sc}\right)$ that
perform the coordinate changes, are obtained. According to
 \eqref{eq:cartSN},
$\phi_{sc}$ ranges in the $X_e$-$Z_e$ plane starting from
the $X_e$ axis. The $(\theta_{e},\phi_{e})$ angles are defined as follows:
\begin{align}
	\phi_e &=
\left\{
\begin{array}{@{}l@{\quad}l@{\quad}l@{\quad}l@{}}
	\,F\left(\theta_{sc},\phi_{sc}\right)
	&\text{if $\theta_{sc}\neq 0, \pi$} &\text{and}
	&\left\{\begin{array}{l}
	     0\le\phi_{sc}<\pi/2\\
	     \text{or } \phi_{sc} > 3\pi/2\\
	\end{array}\right.
	\\
	\begin{array}{l}
	F\left(\theta_{sc},\phi_{sc}\right) -\pi\\
	F\left(\theta_{sc},\phi_{sc}\right) +\pi\\
	\end{array}
	&
	\left.\begin{array}{l}
	\text{if $0<\theta_{sc}<\pi/2$}\\
	\text{if $\pi/2\le\theta_{sc}<\pi$}\\
	\end{array}\right\} & \text{and} &
	\pi/2 < \phi_{sc} < 3\pi/2\\
\end{array}
\right.\\
\theta_e &= 
\tan^{-1}\left(\sin\phi_{sc}/
\sqrt{\cos^2\phi_{sc}+\cot^2\theta_{sc}}\right),
\end{align}
where for $\phi_{sc}=\pi/2, 3\pi/2$ (and $\theta_{sc} = 
\pi/2$) limit may be applied.

\subsection{Solar neutrinos}
\label{app:nuboro}

 
In the case of neutrinos coming from the Sun 
the axis $\hat{d}_{sc}$ (see Fig.~\ref{fig:RFsca}) must point,
time by time, the direction from the source
to the detector.
Since the Earth moves around the Sun, a parametric 
function $f\left(t; \theta_{sc}, \phi_{sc}\right)$ of the time-like
parameter $t$ is needed to reproduce the Earth's orbit when $\theta_{sc} = 0$
and $t\in[0,2\pi)$.\footnote{In order to keep the formulas as simple as
possible, a-dimensional quantities are considered in the following.}

Assuming the Earth orbit as a perfect circle
with the centre in the Sun,
Ref.~\cite{McCabe:2013kea} has been followed.
At order zero,
the Earth's orbit in
standard galactic coordinates can be expressed as:
\begin{equation}\label{eq:ri}
	\hat{r}_i = \cos b_i \cos\left(t - \ell_i\right)
	\quad\text{with}\quad i = x_g,\,y_g,\,z_g,
\end{equation}
where $t \in [0,2\pi)$ is the parameter describing the orbit over one year.
The parameters
$\left(b_i,\ell_i\right)$ needed to describe the axes of the
heliocentric ecliptic rectangular coordinate system
in the galactic one --- see Ref.~\cite{McCabe:2013kea} --- are the following: 
\begin{equation}
	\begin{aligned}
		\left(b_{x,g},\lambda_{x,g}\right) &=
		\left({5}^\circ.536, {266}^\circ.840\right)\\
		\left(b_{y,g},\lambda_{y,g}\right) &=
		\left(-{59}^\circ.574,{347}^\circ.340\right) \\
		\left(b_{z,g},\lambda_{z,g}\right) &=
		\left(-{29}^\circ.811,{180}^\circ.023\right)
	\end{aligned}\qquad\text{with}\quad
	\lambda_i = \ell_i - 180^\circ.
\end{equation}

Let us start with the $\hat{d}_{sc}$ axis placed along the
$X$ axis of a reference frame defined in such a way
that the Earth's orbit lies on the
$X$-$Y$ plane:
\begin{equation}
	\hat{d}_{sc} = (1,\,0,\,0).
\end{equation}
Since the versor $\hat{d}_{sc}$ must rotate according
to the annual Earth motion, 
a time-like parameter $\alpha$ that describes its rotation along the
$Z$-axis has been introduced:
\begin{equation}\label{eq:u1}
	\hat{d}_{1}(\alpha) = \mathcal{R}_{z}(\alpha)\hat{d}_{sc}
	= 
	\begin{pmatrix}
		\cos\alpha 	& -\sin\alpha & 0\\
		\sin\alpha	& \cos\alpha	  & 0\\
		0			& 0			  & 1\\
	\end{pmatrix}
	\begin{pmatrix}
		1\\ 0\\ 0\\
	\end{pmatrix},
\end{equation}
where $\mathcal{R}_{z}$ is the active rotation matrix along the
$Z$ axis.

In order to reproduce the Earth motion in galactic
coordinates \eqref{eq:ri}, the first step is to tilt
the $\hat{d}_1$ orbit by means of a clockwise rotation along
the $X$ axis of an angle $\beta$, according the Eq.\ \eqref{eq:ri}.
Therefore, the passive
$\mathcal{R}^{\mathrm{t}}_x(\beta)$ rotation
matrix to the versor $\hat{d}_{1}\left(\alpha\right)$
defined in Eq.\ \eqref{eq:u1} is applied:
\begin{equation}\label{eq:u11}
	\hat{d}_{2}(\beta,\alpha) = \mathcal{R}^{\mathrm{t}}_x(\beta)
	\hat{d}_{1}\left(\alpha\right)
	=
	\begin{pmatrix}
		1 & 0			& 0\\
		0 & \cos\beta 	& \sin\beta\\
		0 & -\sin\beta	& \cos\beta\\
	\end{pmatrix}
	\begin{pmatrix}
		\cos\alpha 	& -\sin\alpha & 0\\
		\sin\alpha	& \cos\alpha	  & 0\\
		0			& 0			  & 1\\
	\end{pmatrix}
	\begin{pmatrix}
		1\\ 0\\ 0\\
	\end{pmatrix}.
\end{equation}

According the Eq.\ \eqref{eq:u11}, the intersection line between the $X$-$Y$ plane and the tilted
$\hat{d}_{2}$ orbit coincides with the $X$ axis. 
On the other hand,
from Eq.\ \eqref{eq:ri} 
the coordinates $\left(x_g,y_g\right)$ with $z_g=0$ are:
\begin{equation}
	z_g = 0 \Leftrightarrow t = l_z \pm \frac{\pi}{2}
	\quad\Longleftrightarrow\quad
	\begin{cases}
		x_g = \cos b_x \cos\left(l_z-l_x\pm\pi/2\right)
		&\equiv\pm x_{\mathrm{int}}\\
		y_g = \cos b_y \cos\left(l_z-l_y\pm\pi/2\right)
		&\equiv\pm y_{\mathrm{int}}\\
	\end{cases}
\end{equation}
and the azimuth position $\phi_g$ of the intersection point is:
\begin{equation}
	\phi_g = \arctan\left(
	\frac{y_{\mathrm{int}}}{x_{\mathrm{int}}}\right)
	= 0.111419 \equiv \phi_{\mathrm{int}}.
\end{equation}
Therefore, to reproduce the orbit in Eq.\ \eqref{eq:ri} 
the versor $\hat{d}_2\left(\beta,\alpha\right)$ is rotated
by an angle $\gamma = \phi_{\mathrm{int}}$ along the $Z$
axis:
\begin{equation}
\begin{aligned}
	\hat{d}_3\left(\gamma,\beta,\alpha\right) &=
	\mathcal{R}_{z}(\gamma)\hat{d}_2\left(\beta,\alpha\right)\\
	&=
	\begin{pmatrix}
		\cos\gamma 	& -\sin\gamma & 0\\
		\sin\gamma	& \cos\gamma	  & 0\\
		0			& 0			  & 1\\
	\end{pmatrix}
	\begin{pmatrix}
		1 & 0			& 0\\
		0 & \cos\beta 	& \sin\beta\\
		0 & -\sin\beta	& \cos\beta\\
	\end{pmatrix}
	\begin{pmatrix}
		\cos\alpha 	& -\sin\alpha & 0\\
		\sin\alpha	& \cos\alpha	  & 0\\
		0			& 0			  & 1\\
	\end{pmatrix}
	\begin{pmatrix}
		1\\ 0\\ 0\\
	\end{pmatrix}.
\end{aligned}
\end{equation}
Performing the matrix product the $z$-component of  $\hat{d}_3\left(\gamma,\beta,\alpha\right)$ can be obtained as follows:
\begin{equation}
	\hat{d}_{3,z}\left(\gamma,\beta,\alpha\right)
	= -\sin\alpha\,\sin\beta.
\end{equation}
It corresponds to the third component of $\hat{r}$
in Eq.\ \eqref{eq:ri} if one identifies:
\begin{equation}
	\alpha\equiv t-\lambda_z +\frac{\pi}{2}
	\quad\text{and}\quad \beta \equiv \frac{\pi}{2} + b_z.
\end{equation}

Finally, the $\hat{d}_{sc}$ axis can be replaced with a generic
scattering versor $\hat{u}_{sc}$. Since an initial values of 
$\hat{d}_{sc}$ has been assumed along the $x$-axis, a possible parameterisation
could be:
\begin{equation}
	\hat{u}_{sc} = \left(\cos\theta_{sc},
	\sin\theta_{sc}\cos\phi_{sc},
	\sin\theta_{sc}\sin\phi_{sc}\right).
\end{equation}
The final expression is:
\begin{equation}
	\hat{u}_{g}\left(\theta_{sc},\phi_{sc};
	\alpha,\beta,\gamma\right) =
	\mathcal{R}_{z}(\gamma)\mathcal{R}^{\mathrm{t}}_x(\beta)
	\mathcal{R}_{z}(\alpha)\hat{u}_{sc}\left(\theta_{sc},
	\phi_{sc}\right),
\end{equation}
or:
\begin{equation}
	\hat{u}_g=
	\begin{pmatrix}
		\cos\gamma 	& -\sin\gamma & 0\\
		\sin\gamma	& \cos\gamma	  & 0\\
		0			& 0			  & 1\\
	\end{pmatrix}
	\begin{pmatrix}
		1 & 0			& 0\\
		0 & \cos\beta 	& \sin\beta\\
		0 & -\sin\beta	& \cos\beta\\
	\end{pmatrix}
	\begin{pmatrix}
		\cos\alpha 	& -\sin\alpha & 0\\
		\sin\alpha	& \cos\alpha	  & 0\\
		0			& 0			  & 1\\
	\end{pmatrix}
	\begin{pmatrix}
		\cos\theta_{sc}\\
		\sin\theta_{sc}\cos\phi_{sc}\\
		\sin\theta_{sc}\sin\phi_{sc}\\
	\end{pmatrix}.
\end{equation}

The passage from the standard galactic reference frame to the
emulsion one is straightforward:
\begin{equation}
	\hat{u}_e = \left(x_e, y_e, z_e \right) =
	\left(y_g,x_g, -z_g\right)
\end{equation}
and starting from the Cartesian components $x_e$, $y_e$,
$z_e$  the polar angles
$\left(\theta_e,\phi_e\right)$ can be derived 
using the Eq.\ \eqref{eq:fiSN} and \eqref{eq:tetaSN}.


\printbibliography

\end{document}